\documentclass[aps,prd,twocolumn,10pt,superscriptaddress,amssymb,amsmath,nofootinbib]{revtex4-2}

\usepackage[utf8]{inputenc}
\usepackage{lmodern}
\usepackage[T1]{fontenc}
\usepackage{graphicx} 
\usepackage[pdftex,breaklinks,colorlinks,
linkcolor=Blue,
citecolor=teal,
anchorcolor=red,
urlcolor=cyan,
pdfencoding=auto]{hyperref}
\usepackage[dvipsnames]{xcolor}
\usepackage{mathtools}
\usepackage{mathrsfs}
\usepackage{booktabs}
\usepackage{soul}
\usepackage{orcidlink}
\usepackage[mathscr]{euscript}
\newcommand{\beq}{\begin{equation}\begin{aligned}}
\newcommand{\eeq}{\end{aligned}\end{equation}}

\def\Tpp{\mathsf{T}^{(\textrm{p})}}
\def\T{\mathsf{T}}
\def\e{\epsilon}

\def\Op{\mathcal{O}(\epsilon)}
\def\Od{\mathcal{O}(\xi\epsilon)}
\def\F{\mathsf{F}_{\Omega}}
\def\tpp{^{(1,0)}}

\def\tpd{^{(1,1)}}

\def\mfa{\mathfrak{a}}
\def\mfb{\mathfrak{b}}
\def\mfc{\mathfrak{c}}

\newcommand{\Mbh}{\textup{M}_{\textrm{BH}}}
\newcommand{\Mh}{\textup{M}_{\textrm{halo}}}
\newcommand{\rhoF}{\rho_{\textrm{gal}}}
\newcommand{\rhoB}{\rho_{\textrm{BH}}}

\usepackage{hyperref}
\hypersetup{colorlinks=true}

\def\equationautorefname~#1\null{%
	Eq.~(#1)\null
}
\def\figureautorefname~#1\null{%
	Fig.~#1\null
}
\def\tableautorefname~#1\null{%
	Table.~#1\null
}
\def\sectionautorefname~#1\null{%
	Section #1\null
}
\def\appendixautorefname~#1\null{%
	Appendix #1\null
}

\newcommand{\NCUa}{Department of physics, Nanchang University, Nanchang, 330031, China}
\newcommand{\NCUb}{Center for Relativistic Astrophysics and High Energy Physics, Nanchang University, Nanchang, 330031, China}

\begin{document}

\title{
Probing beyond-vacuum general relativistic effects with extreme mass-ratio inspirals}

\author{Tieguang Zi\,\orcidlink{0000-0003-0046-2056}}
\affiliation{\NCUa}
\affiliation{\NCUb}

\author{Mostafizur Rahman \orcidlink{0000-0003-0904-8548}}
\affiliation{Department of Physics, Kyoto University, Kyoto 606-8502, Japan}

\author{Shailesh Kumar\,\orcidlink{0000-0001-7072-9452}}
\affiliation{Indian Institute of Technology, Gandhinagar, Gujarat-382355, India}

\begin{abstract}
We examine extreme mass-ratio inspirals (EMRIs) as probes of beyond-vacuum general relativistic effects, accounting for both astrophysical environments and scalar Gauss-Bonnet (sGB) gravity. In beyond-vacuum scenarios, the evolution of an EMRI immersed in a cold dark matter environment modifies the gravitational wave flux and introduces additional dissipative effects such as dynamical friction. In parallel, in the beyond-general relativistic settings such as in sGB gravity, the inspiraling object carries an effective scalar charge and emits scalar radiation. Both environmental and modified-gravity effects modify the flux-balance law, thereby inducing changes in the EMRI dynamics. Using a two-timescale analysis within the fixed-frequency formalism, we compute leading-order corrections to the energy fluxes for quasi-circular, equatorial orbits in static, spherically symmetric spacetimes and construct the corresponding gravitational waveforms, which are used to quantify the accumulated gravitational wave dephasing and waveform mismatch relative to the vacuum general relativistic case. We further perform the Fisher Information Matrix analysis to estimate parameter correlations and the ability of future space-based detectors such as the Laser Interferometer Space Antenna (LISA) to disentangle environmental and modified gravity effects. Our results show that both dark matter and scalar field effects can leave measurable imprints on EMRI waveforms and that a consistent beyond-vacuum  treatment is essential for robust tests of gravity.
\end{abstract}

\maketitle

\tableofcontents

\section{Introduction}
As gravitational wave (GW) astronomy enters a precision era \cite{LIGOScientific:2016aoc, LIGOScientific:2017bnn, LIGOScientific:2018dkp, LIGOScientific:2016sjg}, understanding how astrophysical environments and potential physics beyond general relativity (GR), altering GW signals from black hole binaries, has become crucial for correctly disentangling various astrophysical signatures/parameters in the next-generation GW measurements \cite{Gair:2012nm, Berti:2015itd, Amaro-Seoane:2014ela, Cardenas-Avendano:2024mqp, Cardoso:2019rvt}. In realistic astrophysical settings, black holes are expected to reside in non-vacuum environments, where interplay with surroundings like dark matter distributions and additional fields can give rise to detectable modifications in gravitational waveforms \cite{Cardoso:2021wlq, Barausse:2014pra}. If ignored, such effects can bias parameter estimation, producing spurious implications of non-GR imprints and further suppressing the correct information of strong gravity regions; whereas, incorporating them consistently allows a more precise evaluation of beyond-vacuum GR effects that can possibly bring improvements on the associated constraints, which can accurately be detected by the future space-based detectors \cite{Yunes:2009ke, LISA:2022kgy, Barausse:2014tra, Kocsis:2011dr, Romero-Shaw:2021ual, Lyu:2024gnk}. This requirement becomes particularly essential as detectors achieve unprecedented sensitivities and begin observing extreme mass-ratio inspirals (EMRIs)--one of the most informative and promising GW sources--whose long-lasting signals accumulate phase over thousands to millions of orbital cycles \cite{Amaro-Seoane:2014ela, Cardenas-Avendano:2024mqp, Babak:2017tow}. An EMRI is a two-body system where a stellar-mass compact object (the secondary) inspirals a supermassive black hole (the primary), acting as a lens for subtle and minute observable corrections, that makes the system exceptionally sensitive to environmental influences and potential beyond-GR footprints. As missions like Laser Interferometer Space Antenna (LISA) \cite{amaroseoane2017}, TianQin \cite{TianQin:2015yph, TianQin:2020hid}, Taiji \cite{Ruan:2018tsw}, 
and DECIGO~\cite{Seto:2001qf, Kawamura:2020pcg} move toward operation, developing EMRI waveform models that capture beyond-vacuum GR imprints is therefore crucial, both for high-precision tests of gravity and for determining whether environmental or non-GR signatures can be disentangled from standard vacuum EMRIs dynamics, consequently ensuring robust and accurate inference of physics in the strong gravity fields~\cite{Sadeghian:2013laa,Zi:2021pdp,Zi:2023omh,Rahman:2021eay,Kumar:2024utz,Rahman:2023sof,Zi:2025jxy,Speeney:2024mas,Speri:2022upm,Duque:2023seg,Cardoso:2022whc,Zi:2024itp,Figueiredo:2023gas,Rahman:2025mip}.

Of these considerations, we understand that supermassive black holes are expected to reside commonly in dense galactic environments, where surrounding matter can alter the inspiral dynamics relative to the standard vacuum case~\cite{Barausse:2014tra,Copparoni:2025vty,Cheng:2025wac}.
If dark matter is composed of cold, collisionless particles, the slow (adiabatic) growth of a black hole can produce a pronounced overdensity region--commonly referred to as a ``spike''--around supermassive, intermediate--mass, and even primordial black holes~\cite{Kavanagh:2020cfn, Gondolo:1999ef,Sadeghian:2013laa,Ullio:2001fb,Eda:2013gg,Guo:2017njn,Barsanti:2021ydd,Kuhnel:2018mlr,Speeney:2022ryg,Jiang:2023xwv,Tahelyani:2024cvk,Cheng:2024mgl,Li:2025qtb,Zhang:2024ugv,Mitra:2025tag,Karydas:2025gqp,Gliorio:2025cbh,Wade:2025rkk,Karmakar:2025drp,Wang:2025msq}.
Such overdensities can, in principle, be inferred by measuring the detectable changes they imprint on the GW signal from compact objects orbiting the black hole~\cite{Eda:2013mva,Eda:2015cza,Wade:2025rkk}.
As the compact object orbits in such an environment, it experiences a gravitational drag from the surrounding medium, known as dynamical friction~\cite{Chandrasekhar:1943zz,Kim:2007zb,Traykova:2021dua,Dosopoulou:2023umg,Traykova:2023qyv,Cole:2022ucw,Buehler:2022tmr,Spieksma:2025exm,Li:2025qtb,Cheng:2025wac},
which affects the rates of change of orbital energy and angular momentum and introduces a dissipative contribution that can noticeably modify the long-term orbital evolution of EMRIs, thereby inducing waveform dephasings~\cite{Barausse:2014tra,Eda:2013mva,Eda:2015cza,Destounis:2022obl,Li:2025ffh,Wang:2025msq,Duque:2024mfw,Khalvati:2024tzz,Spieksma:2024voy,Duque:2025yfm}.
There are multiple indirect evidences available, supporting the existence of dark matter \cite{Bertone:2004pz, Navarro:1995iw, Freese:2008cz, Clowe:2006eq, Persic:1995ru, Corbelli:1999af, Massey:2010hh, Ellis:2010kf},
and its interaction with EMRIs offers a promising route to probe subtle and potentially observable properties of this elusive component of the Universe~\cite{Macedo:2013qea, Cardoso:2021wlq, Figueiredo:2023gas, Cardoso:2022whc, Barausse:2014pra, Rahman:2023sof, Rahman:2025mip, Eda:2013mva,Eda:2015cza}.

On the other hand, among the best-studied extensions to GR relevant for compact-object systems are models in which a scalar field couples directly to curvature invariants, most notably the Gauss-Bonnet term (scalar Gauss-Bonnet theory) \cite{Sotiriou:2013qea, Julie:2019sab}.
In this class of theories, broad ``no-hair'' considerations often imply that, for EMRIs modelling purposes, the central massive black hole is well approximated by the Kerr metric at leading order, while the compact secondary generically carries a (small) scalar charge and sources scalar radiation during the inspiral~\cite{Maselli:2020zgv,Barsanti:2022gjv,Maselli:2021men, Barsanti:2022ana, DellaRocca:2024pnm,Spiers:2023cva,Zi:2025lio, Speri:2024qak}.
The cumulative impact of such scalar emission over the long EMRIs evolution can produce measurable deviations in the waveforms, enabling stringent constraints on the scalar charge and, when applicable, on the scalar-field mass~\cite{Maselli:2020zgv,Barsanti:2022gjv,Xie:2024xex,DellaRocca:2024sda}.
Related approaches model the inspiralling object in a skeletonized manner as an effective point particle endowed with a scalar-field--dependent mass function~\cite{Eardley:1975zz,Damour:1992we,Damour:1996ke}.
Therefore, disentangling environmental imprints from beyond-GR effects such as scalar charges is crucial for correctly interpreting EMRI signals and securing robust constraints on new physics beyond-vacuum scenarios~\cite{Barausse:2014tra,Maselli:2020zgv,Barsanti:2022gjv,LaHaye:2025ley,Li:2025ffh}.

Significant advancements have been made in developing high-accuracy vacuum EMRIs waveforms and evolution schemes, including first-order self-force ingredients, spin effects, and systematic multiscale (two-timescale) formalisms for long-duration inspirals \cite{Barack:2018yvs,Hinderer:2008dm,Miller:2020uly, Mathews:2021rod,Mathews:2021rod, Pound:2021qin,PhysRevLett.109.051101, PhysRevD.92.104047}. In parallel, a broad class of practical waveform constructions relies on frequency-domain, fixed-frequency building blocks for Kerr (or Schwarzschild) geodesic motion, which provide the infrastructure for assembling accurate fluxes and waveform modes along an adiabatically evolving inspiral \cite{Hughes:1999bq,Fujita:2004bv,Khalvati:2025znb}. 
However, EMRIs are generically expected to form and evolve in environments where the spacetime is not strictly vacuum. Astrophysical matter distributions and additional degrees of freedom from extensions of GR can introduce further perturbations--through changes to the effective background, new channels of radiation, and dissipative drags--that modify both the inspiral rate and the accumulated phase of the signal \cite{Barausse:2014tra,Zwick:2025ine,Aurrekoetxea:2024cqd,Alnasheet:2025mtr}. Of particular interest are scenarios in which the primary is embedded in a dense dark matter configuration (e.g., cusps/spikes produced by adiabatic black-hole growth) \cite{Gondolo:1999ef,Bertone:2024dmwg}, and cases in which the secondary carries a small scalar charge, as occurs in sGB-type theories \cite{Maselli:2020zgv,Barsanti:2022gjv}. In these settings, scalar radiation, matter-induced modifications to the dynamics, and environmental dissipative effects (such as dynamical friction) can enter at the same perturbative order as the conventional self-force corrections, and may therefore produce comparable imprints on the long-term phasing \cite{Burke:2023lno,Gliorio:2025cbh,Nicolini:2025yhu}. 
Consequently, incorporating scalar-field effects and environmental corrections within two-timescale evolution—while retaining fixed-frequency, mode-by-mode waveform building blocks—provides a unified framework to assess (i) how non-vacuum physics leaves detectable features in EMRIs waveforms and (ii) whether LISA can disentangle beyond-GR scalar effects from astrophysical environmental imprints \cite{Hinderer:2008dm,Wilcox:2024sqs,Miller:2020uly,Barsanti:2022gjv,Cardoso:2023dwz,Kugarajh:2025rbt}.

Moving beyond the idealized vacuum-GR scenario, the modification of the Einstein field equation can be schematically represented as
\beq
G_{\mu\nu}=8\pi \left(\Tpp_{\mu\nu}+\alpha_{\textrm{bGR}} \T_{\mu\nu}^{\textrm{bGR}}\right)
\eeq
where $G_{\mu\nu}$ is the Einstein tensor, $\Tpp_{\mu\nu}$ is the point-particle's energy momentum tensor, and the parameter $\alpha_{\textrm{bGR}}$ characterizes beyond-vacuum GR correction terms, which are collectively represented here as $T_{\mu\nu}^{\textrm{bGR}}$. In this paper, $T_{\mu\nu}^{\textrm{bGR}}$ represents the beyond-vacuum GR correction terms, which we consider to arise either due to the presence of dark matter or due to the presence of additional degrees of freedom in the form of a scalar field.\par 
Departing from the idealized vacuum-GR framework introduces additional channels through which the secondary object can lose energy and angular momentum. As a result, the flux balance law is modified. For a secondary object inspiraling around a supermassive black hole on a quasicircular orbit, this modification can be schematically represented as
\beq\label{flux_balance_law_0}
-\dot{\textup{E}}_{\textrm{orb}} &= \dot{\textup{E}}_{\textrm{GR}}+\mathcal{P}(\alpha_{\textrm{bGR}})\dot{\textup{E}}_{\textrm{bGR}}~,\\
\eeq
where $-\dot{\textup{E}}_{\textrm{orb}}$ represents the rate at which the secondary its loss of orbital energy, $\dot{\textup{E}}_{\textrm{GR}}$ represents the GW flux in vacuum GR scenario, and $\mathcal{P}(\alpha_{\textrm{bGR}}) \dot{\textup{E}}_{\textrm{bGR}}$ represents the modification of the flux balance law with the parameter $\mathcal{P}(\alpha_{\textrm{bGR}})$ characterizes this modification.  We adopt \textit{fixed-frequency} formalism \cite{Mathews:2021rod, Mathews:2025nyb} to study this problem and write the flux balance law in the form described in \autoref{flux_balance_law_0}. 

Further, in this line of endeavour, recent works on probing the beyond-vacuum case have shown that environmental modifications to GR can leave detectable imprints on EMRIs~\cite{Kejriwal:2023djc, Bhalla:2024lta,Zwick:2024yzh,Kejriwal:2025upp,Kejriwal:2025jao,Yuan:2024duo,Lyu:2025zql,Copparoni:2025jhq,Dosopoulou:2025jth}, while being strongly correlated with standard source parameters in Fisher-matrix analyses. Using planetary migration (PM) effect due to the accretion disk surrounding the massive black hole and a time-varying gravitational constant as representative examples~\cite{Kejriwal:2023djc, Kejriwal:2025upp,Kejriwal:2025jao, Yuan:2024duo}, it has been demonstrated that neglecting such contributions can bias inference \cite{Kejriwal:2023djc}, whereas bias-correction (based on bias-corrected importance sampling technique) and population-level analyses can disentangle competing beyond-vacuum scenarios \cite{Kejriwal:2025upp, Kejriwal:2025jao}. These studies also motivate a systematic and explicit investigation of the combined effects. In the present work, we investigate EMRI dynamics in the presence of distinct dark matter profiles and beyond-GR effects, and further, building on the two-timescale and self-force formalisms, we consistently incorporate both dark matter–induced modifications of the background geometry and scalar-field–driven perturbations from sGB theory into the fluxes, waveform, and orbital phase evolution. This cohesive framework assesses their combined impact on waveform detectability, parameter correlations, and measurement uncertainties using the Fisher Information Matrix (FIM). Therefore, it sets up a framework to characterize, within a consistent perturbative scheme, the extent to which beyond-vacuum physics can imprint measurable signatures on EMRI signals, to examine potential degeneracies among environmental and modified gravity effects/parameters, and to assess the related constraints achievable with LISA observations. Also, this work provides an order of magnitude estimate for explicit treatment of beyond-vacuum GR effects, providing an initial quantitative analysis that lays the groundwork for precision waveform modelling in strong gravity fields.

We now briefly outline the structure of the paper. In  \autoref{sec2}, we present the theoretical framework, describing the spherically symmetric background spacetime and the modelling of EMRIs via perturbation equations and fluxes in astrophysical environments characterized by dark matter density profiles. In \autoref{secIII}, we introduce the sGB theory adopted in this work, outline the underlying assumptions, and structure the perturbative equations and flux contributions. In \autoref{secIV}, we provide the details regarding estimation on beyond-vacuum parameters through waveform generation, FIM, dephasing and mismatch analysis. \autoref{result} is devoted to the results on detectability, parameter-estimation uncertainties and correlations induced by environmental and modified gravity effects. Finally, we summarize our results and discuss future directions in \autoref{secVI}.
\section{EMRI system in the presence of dark matter}\label{sec2}
\subsection{Static, spherically symmetric black hole immersed in dark matter}
 In this section, we describe an EMRI system immersed in a cold dark matter environment.  In this work, we consider two dark matter models, namely the Hernquist profile \cite{1990ApJ...356..359H} and the Navarro-Frenk-White (NFW) profile \cite{Navarro:1996gj}. The dark matter density for these profiles can be expressed as
 \begin{equation}\label{density_galaxy}
    \begin{aligned}     \rhoF(r)=\rho_0\left(\frac{r}{a_0}\right)^{-\gamma}\left[1+\left(\frac{r}{a_0}\right)^\alpha\right]^\frac{\gamma-\beta}{\alpha}~.
    \end{aligned}
\end{equation}
In the above equation, $a_0$ is the scale radius, and $\rho_0$ is the scale density which is proportional to the dark matter density at the scale radius, $\rho_0=2^{(\beta-\gamma)/\alpha}\rhoF(a_0)$. Furthermore, the parameters $(\alpha,\beta,\gamma)$ take the value $(1,4,1)$ and $(1,3,1)$ for the Hernquist and NFW dark matter profiles, respectively.
 \par 
In Ref.~\cite{Sadeghian:2013laa}, the authors studied the adiabatic growth of a Schwarzschild black hole inside dark matter halo. They demonstrated that when the black hole grows at the center of a dark matter halo, its gravitational pull redistributes the surrounding dark matter particles which leads to the formation of a so-called \textit{dark matter spike} profile. The  density of dark matter spike profile can be written in the following manner ~\cite{Sadeghian:2013laa}
 \begin{align}\label{rho}
    \rho_{\rm BH}(r)=\frac{4\pi}{r^2\sqrt{f}}
   \int \frac{{\cal E}LF( E({\cal E},L))d{\cal E}dL}{\sqrt{{\cal E}^2-f(1+L^2/r^2)}}~.
\end{align}
where $\Mbh$ is the mass of the final black hole, $f = 1 - 2\Mbh/r$, and $F(E)$ denotes the phase-space distribution function, with $E$ representing the initial energy of the dark matter halo. This phase-space distribution function can be obtained by inverting Eddington’s inversion formula~\cite{Lacroix:2018qqh}. Since the central black hole grows through an adiabatic process, the adiabatic invariants remain unchanged before and after the growth of the black hole ~\cite{Sadeghian:2013laa}. By equating the azimuthal and polar adiabatic invariants, we can find that the angular momentum of the system remains unchanged between the initial and final states. Furthermore, by equating the radial adiabatic invariant, the initial-state energy $E$ of dark matter halo can be expressed as a function of the final-state energy $\mathcal{E}$ and the angular momentum of the spike profile, i.e., $E=E(\mathcal{E},L)$.  Notably for Hernquist and NFW profiles, the density of the dark matter spike profile can be written as \cite{Speeney:2024mas, Rahman:2025mip}
\begin{equation}\label{density_Cardoso}
    \begin{aligned}
   \rhoB(r)
   =\Mh\frac{\bar{\rho}_s(r)}{I(r_c)}\Theta(r-4\Mbh),
    \end{aligned}
\end{equation}
where $\Mh$ is total mass of the dark matter profile, $r_c$ represents the cut off radius of the spike profile with $\rhoB(r)=0$ for $r>r_c$, $\Theta(r-4\Mbh)$ is the Heaviside step function, and  
\begin{equation}\label{density_Cardoso1}
    \begin{aligned}
   \bar{\rho}_s(r)
   &=\left(1-\frac{4 \Mbh}{r}\right)^{ \mfa}\left(\frac{R_S}{r}\right)^{\mfb}\left(1+\frac{r}{R_S}\right)^{-\mfc}, \\
    I(r)
   &=\int_{4\Mbh}^{r}4\pi r^2 \bar{\rho}_s(r) dr.
    \end{aligned}
\end{equation}
 Here, $R_S=\Mh a_0/\Mbh$. The  three numerical parameters $(\mfa$,$\mfb$,$\mfc)$ are obtained by numerically integrating \autoref{rho} and then fitting with the profile described in \autoref{density_Cardoso} and \autoref{density_Cardoso1}. 
 Following \cite{Rahman:2025mip}, we consider a dark matter spike profile with $\Mh = 10^4 \Mbh$, $\Mh/a_0 = 10^{-3}$, and $r_{\rm c} = 100 \Mh a_0 / \Mbh$. The numerical fitting parameters for the Hernquist and NFW profile in that case is presented in \autoref{tab:Fitting_Parameters}.  \par
\begin{figure*}[t!]
\centering	
\includegraphics[width=3.2in, height=2.2in]{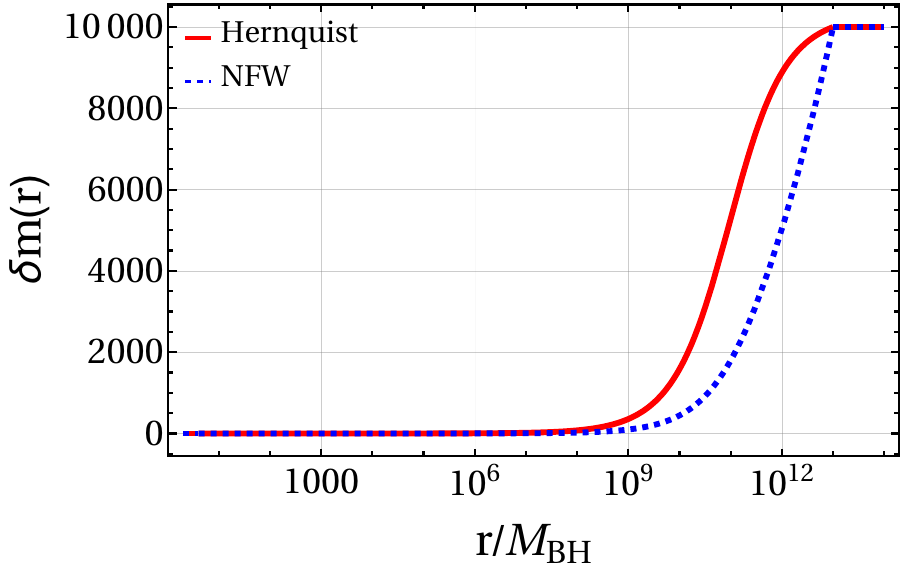}
\includegraphics[width=3.2in, height=2.12in]{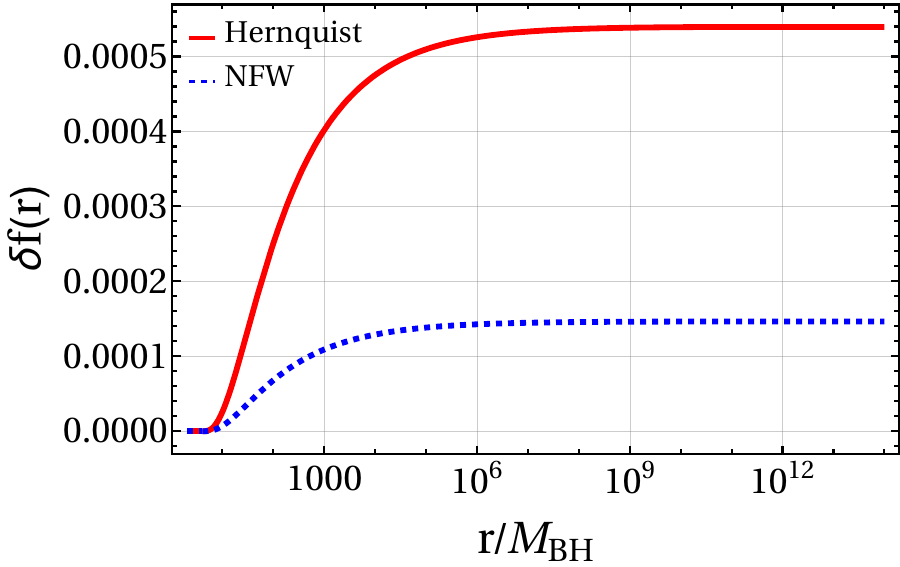}
\caption{The plot of the $\delta m(r)$ (left panel) and  $\delta f(r)$ (right panel) as the functions of $r$. In each of these plots the red curve represents the Hernquist spike profile whereas the blue curve represents the NFW spike profile. The mass of the spike profile is taken to be $\Mh=10^4\Mbh$  with $\Mh/a_0=10^{-3}$  and the cut off radius $r_c=100 \Mh a_0/\Mbh$. The fitting parameters $\{\mfa,\mfb,\mfc\}$ are the same as \autoref{tab:Fitting_Parameters}.}\label{fig: radial_shift}
\end{figure*}

\begin{table}[t!]
\centering
 \def\arraystretch{1.1}      	
	\setlength{\tabcolsep}{1.2em}
\begin{tabular}{|l|l|l|l|}
\hline\hline
Fitting Parameters      & $\mfa$ & $\mfb$   & $\mfc$                                                                       \\\hline\hline

Hernquist        & 2.637                     & 2.330                        & 1.344                                                                                            \\
NFW              & 2.640                    & 2.332                        & 0.5446                                                                                          
\\\hline\hline
\end{tabular}
\caption{The numerical values of the fitting parameters $(\mfa\,,\mfb\,,\mfc)$ for the Hernquist and NFW spike profiles, as defined in \autoref{density_Cardoso1}, are presented. The fits are obtained by assuming a total dark matter halo mass of $\Mh = 10^{4}\Mbh$ and adopting the scale radius $a_{0} = 10^{3}\Mh$. Furthermore, the cutoff radius is taken to be $r_{\rm c} = 100 \Mh a_0 / \Mbh$.
}\label{tab:Fitting_Parameters}
\end{table}
To describe a static and spherically symmetric spacetime in the presence of a dark matter spike, Ref.~\cite{Rahman:2025mip} adopted a perturbative approach. The resulting spacetime can be written as
\beq\label{DM_spacetime}
g_{\mu\nu}=\bar{g}_{\mu\nu}+\xi h^{(0,1)}_{\mu\nu}+\mathcal{O}(\xi^2)~.
\eeq
where $\bar{g}_{\mu\nu}$ denotes the background Schwarzschild spacetime, and $h^{(0,1)}_{\mu\nu}$ encodes the leading-order modification of the background geometry induced by the dark matter distribution. The parameter $\xi$ characterizes the order of this modification. We refer this parameter as the \textit{dark matter parameter}. Since we are interested only in the leading-order corrections to the GW flux arising from the presence of dark matter, it is sufficient for our purposes to consider the spacetime metric in the form given in \autoref{DM_spacetime}. Furthermore, describing the dark matter as an anisotropic fluid with vanishing radial pressure, i.e., $\T^{\textrm{(0,1)}\mu}{}_\nu=\textrm{diag}(\rhoB,0,p_t,p_t)$ \cite{PhysRevD.105.L061501, PhysRevD.67.104017}, we can find the non vanishing components of $h^{(0,1)}_{\mu\nu}$ by solving the Einstein field equation $G_{\mu\nu}=8\pi \xi \T^{\textrm{(0,1)}}_{\mu\nu}$ up to order $\mathcal{\xi}$, which are given by 
\begin{equation}\label{Gtt1p}
    \begin{aligned}
    h^{(0,1)}_{rr}=\frac{2\delta m(r)}{f^2 r}\,,\quad
     h^{(0,1)}_{tt}=f\delta f(r)-\frac{2\delta m(r)}{r}~, \\
    \end{aligned}
\end{equation}
where we refer $\delta m(r)$ as the mass function and $\delta f (r)$ as the red-shift function. The expressions for the mass function and red-shift function along with the expression of tangential pressure $p_t$ can be obtained from the dark matter density $\rhoB$ through the relation 
\begin{equation}\label{Gtt1pd}
\begin{aligned}
\delta m'(r)=4\pi r^2\rhoB\,,\quad
\delta f'(r)= \frac{r\rhoB}{2\pi f}\,,\quad
p_t &=\frac{\Mbh \rhoB }{2 r f }~.
\end{aligned}
\end{equation}
where ``prime'' denotes derivative of the function with respect to its argument. 
In \autoref{fig: radial_shift}, we plot the mass function $\delta m(r)$ and the radial metric function $\delta f(r)$ as functions of the radial coordinate for the Hernquist and NFW dark matter spike profiles. As in \autoref{tab:Fitting_Parameters}, we consider a dark matter spike with total mass $\Mh = 10^4 \Mbh$, $\Mh/a_0 = 10^{-3}$, and a cutoff radius $r_{\rm c} = 100 \Mh a_0 / \Mbh$. The fitting parameters ${\mfa, \mfb, \mfc}$ are the same as those presented in \autoref{tab:Fitting_Parameters}. We note that the mass function vanishes for $r < 4\Mbh$ and asymptotically approaches $\Mh$ near the cutoff radius. Similarly, the redshift function vanishes for $r < 4\Mbh$ and gradually settles to a small value of order $\mathcal{O}(10^{-4})$ near the cutoff radius.\par
Note that, the Misner-Sharp mass of the black hole-dark matter system is given by $M_{\textrm{MS}}=\Mbh+\xi \delta m$. Finally, the justification for considering the problem in the perturbative approach as descibed in \autoref{DM_spacetime} stems from the fact that for the configuration described above, the dark matter parameter takes the value
\beq\label{delta_m}
\xi=\left(\frac{\Mh}{\Mbh}\right)\frac{1}{L\left[y(r_c)\right]}\approx\left(\frac{\Mh}{\Mbh}\right)\left(\frac{4\Mbh}{r_c}\right)^{3-\mfb}~.
\eeq
where $y(r)=1-4\Mbh/r$ and the function $L[y]$ is given by 
\begin{equation} \label{density}
    \begin{aligned}
    L[y]=\frac{y^{\mfa+1}}{\mfa+1}F_{1}\left(\mfa+1,-(\mfb+\mfc-4),\mfc,\mfa+2,y,\zeta y\right)~,
    \end{aligned}
\end{equation}
 with $\zeta=R_S/(R_S+4\Mbh)$, and   $F_1(\alpha,\beta,\gamma,\delta,y,\zeta y)$ is the Appell hypergeometric function \cite{nakagawa2023appelllauricella}. Given that for any realistic astrophysical situation $r_c\gg \Mbh$ and fitting parameter $\mfb<3$ (see \autoref{tab:Fitting_Parameters}), the dark matter parameter value is much smaller than unity. Specifically,  
for the configuration presented in \autoref{tab:Fitting_Parameters}, the dark matter parameter takes the value $\xi = 5.65 \times 10^{-4}$ ($\xi = 1.53 \times 10^{-4}$) for the Hernquist (NFW) spike profile.
\subsection{EMRI system in the presence of dark matter}
In this section, we consider an EMRI system embedded in a dark matter environment. The evolution of such an EMRI system was studied by one of the authors in Ref.~\cite{Rahman:2025mip}. Here, we summarize the main results of that work and refer interested readers to Ref.~\cite{Rahman:2025mip} for detailed calculations. We assume that the spacetime metric of the primary black hole in the presence of dark matter is given by \autoref{DM_spacetime}. The secondary object induces perturbations both to the spacetime geometry and to the fluid configuration described in the previous section. Consequently, the spacetime and the fluid energy momentum tensor in the presence of a secondary of mass $m_p$ can be described as ~\cite{Dyson:2025dlj, Rahman:2025mip}
\beq\label{metric_emri}
\mathbf{g}_{\mu\nu}&=\bar{g}_{\mu\nu}+h_{\mu\nu} ~,\\
\textup{T}^{\textrm{m}}_{\mu\nu}&=\textup{T}^{\textrm{m}}_{\mu\nu}[\bar{g}+h]=\xi~\T^{\textrm{m(0,1)}}_{\mu\nu}+\e\xi ~\T^{m\textrm{(1,1)}}_{\mu\nu}~, \\
h_{\mu\nu} &=\xi h^{(0,1)}_{\mu\nu}+\epsilon h^{(1,0)}_{\mu\nu}+\epsilon \xi h^{(1,1)}_{\mu\nu}~,
\eeq
where $\epsilon$ counts the power of mass ratio $q$. Here, we ignore the higher order $\mathcal{O}(\xi^2,\epsilon^2)$ contributions to the metric perturbation. In the above expression and in elsewhere of the paper, we assume that the terms with bracketed super and subscript like $h^{(n,k)}_{\mu\nu}$ and $\T^{m\textrm{(n,k)}}_{\mu\nu}$ appears at order $\epsilon^n \xi^k$. The energy momentum tensor of the secondary object is 
\beq\label{pp_em}
\textup{T}^{\textrm{p}}_{\mu\nu}[\mathbf{g}] = m_p \int d\tau \frac{\delta^{(4)}(x-z(\tau))}{\sqrt{-\mathbf{g}}} {u}_{\mu} {u}_{\nu},
\eeq
where $z^{\mu}$ is the worldline of the object and  ${u}^{\mu}$ denotes its four-velocity. Replacing \autoref{metric_emri} in Einstein field equation $G_{\mu\nu}[\mathbf{g}]=8\pi (\textup{T}^{\textrm{m}}_{\mu\nu}[\mathbf{g}]+\textup{T}^{\textrm{p}}_{\mu\nu}[\mathbf{g}])$ and expanding in power of $\e$ and $\xi$, we obtain 
\begin{align}
\delta G^{(0,1)}_{\mu\nu}&=8\pi \T^{\textrm{m(0,1)}}_{\mu\nu}~,\label{mat}\\
\delta G^{(1,0)}_{\mu\nu}&=8\pi \T^{\textrm{p(1,0)}}_{\mu\nu}~,\label{pp}\\
\delta G^{(1,1)}_{\mu\nu}&=8\pi \T^{\textrm{p(1,1)}}_{\mu\nu}+8\pi \T^{\textrm{m(1,1)}}_{\mu\nu} -\delta^2 G^{(1,1)}_{\mu\nu}~,\label{matpp}
\end{align}
where $\delta G^{(n,k)}_{\mu\nu}$ ($\delta^2 G^{(n,k)}_{\mu\nu}$) denotes the linear (quadratic) contribution to the Einstein tensor corresponding to the coefficient of $\epsilon^n \xi^k$ (see Eq.~(29) of \cite{PhysRevD.103.064048}). Similarly, $\mathsf{T}^{\textrm{m}(n,k)}_{\mu\nu}$ ($\mathsf{T}^{\textrm{p}(n,k)}_{\mu\nu}$) represents the fluid (point-particle) energy-momentum tensor component at order $\epsilon^n \xi^k$. Furthermore, the fluid energy-momentum tensor satisfies the covariant conservation equation, which, when expressed with respect to the background spacetime, takes the following form
\begin{align}
\nabla_{\nu}\mathsf{T}_{\textrm{m(0,1)}}^{\mu\nu}&=0\,,\label{cov_cons_law01}\\
 \nabla_{\nu}\mathsf{T}_{\textrm{m(1,1)}}^{\mu\nu}&+C^{\mu}_{\nu\alpha}[h^{(1,0)}]\mathsf{T}_{\textrm{m(0,1)}}^{\alpha\nu}+C^{\nu}_{\nu\alpha}[h^{(1,0)}]\mathsf{T}_{\textrm{m(0,1)}}^{\mu\alpha}=0\,,\label{cov_cons_law11}
\end{align}
where  $C^{\mu}_{\nu\alpha}[h]=g^{\mu\lambda}(2h_{\lambda(\nu;\alpha)}-h_{\nu\alpha;\lambda})/2$, and $\nabla_{\nu}$ and the symbol “$;$” denote covariant differentiation with respect to the Schwarzschild spacetime. Note that the solution of \autoref{mat}, together with \autoref{cov_cons_law01}, represents the static and spherically symmetric spacetime solution of a black hole immersed in a dark matter environment, as presented in the previous section. Equation \autoref{pp} represents the Einstein equation for an EMRIs system in vacuum spacetime. Finally, \autoref{matpp} and \autoref{cov_cons_law11} account for the modifications to the Einstein field equations and the covariant conservation equation for EMRIs in a dark matter environment. \par
As a result of the perturbation, the secondary object moves along a timelike geodesic of an effective metric $\hat{g}_{\mu\nu}=\bar{g}_{\mu\nu}+{h}^R_{\mu\nu}$ where ${h}^R_{\mu\nu}$ is obtained by removing certain singular pieces from physical metric perturbation $h_{\mu\nu}$. When expressed in terms of the background metric, the motion of the secondary corresponds to that of a (self-)accelerated object, which follows a forced geodesic equation
\cite{ Mino:1996nk, Quinn:1996am, PhysRevD.103.064048,PhysRevLett.109.051101, PhysRevD.95.104056}
\beq\label{sf_eqn}
u^{\mu}_{;\nu}u^{\nu}&=F^{\nu}_{G}[{h}^R]~, \\ 
&\equiv -\frac{1}{2}P^{\mu\nu}(g_{\nu}^{\lambda}-{h}_{\nu}^{R\lambda})\left(2 {h}^R_{\lambda \rho ; \sigma}-h^R_{\rho \sigma ; \lambda}\right) u^{\rho} u^{\sigma}~,  \\
&=\xi F_{G(0,1)}^{\mu}+\e F_{G(1,0)}^{\mu}+\e \xi F_{G(1,1)}^{\mu}~,
\eeq
where $u^{\mu}$ is the four-velocity in Schwarzschild spacetime, $P^{\mu}_{\nu}=\bar{g}^{\mu}_{\nu}+u^{\mu}u_{\nu}$, and $F^{\mu}_{(n,k)}$ is the component of $F^{\mu}$ at order $\epsilon^n \xi^k$. 
Note that the leading-order force term at order $\mathcal{O}(\xi)$, $F_{G(0,1)}^{\mu}$, leads to a shifted trajectory (with respect to the background spacetime) of a point test-mass object due to the presence of dark matter (\autoref{DM_spacetime}) and is purely conservative,i.e., it modifies the orbital energy and angular momentum of the secondary object. 
In contrast, $F_{G(1,0)}^{\mu}$ and $F_{G(1,1)}^{\mu}$ contain both dissipative and conservative components. Since this paper focuses on the leading-order modifications to the GW flux due to the presence of dark matter, we focus only the dissipative components of these force terms. 
\subsection{Orbital evolution in the presence of dark matter}
In this paper, we assume that the secondary object follows a quasi-circular orbit. To describe the shifted trajectory, we adopt the \textit{fixed-frequency formalism} \cite{Mathews:2021rod}. Within this framework, the orbital frequency of the shifted orbit remains the same as that of the unperturbed orbit, while the orbital radius is shifted \cite{Mathews:2021rod}. Specifically,
\beq\label{r_expand}
\Omega = \Omega_0\,,\quad
r = r_\Omega + \xi r_{(0,1)}(\Omega) + \epsilon r_{(1,0)}(\Omega) + \epsilon\xi  r_{(1,1)}(\Omega),\quad
\eeq
where $\Omega_0$ is the orbital frequency of a test particle in a circular orbit around a Schwarzschild black hole, and
\beq
r_{\Omega} &= \frac{\Mbh}{x}\,,\quad r_{(A)}=-\frac{F^{r}_{(A)}(1-3x)}{3\Omega^2 f_{\Omega}}~, \\
r_{(1,1)}&=-\frac{F^{r}_{(1,1)}(1-3x)}{3\Omega^2 f_\Omega}+\frac{2 r_{(0,1)}r_{(1,0)}x(1-4x)}{\Mbh f_\Omega}~,
\eeq
where $x=(\Mbh\Omega)^{2/3}$, $f_{\Omega}=f(r_{\Omega})$ and $(A)\in \{(0,1),(1,0)\}$. Using \autoref{r_expand}, we can obtain the expression for orbital energy for the secondary object in the presence of dark matter which is given by  
\beq\label{orbital_energy}
\textup{E}_{\textrm{orb}}&=\textup{E}^{(0)}_{\textrm{orb}}+\xi \textup{E}^{(0,1)}_{\textrm{orb}}~, \\
&=u^{0}f_\Omega-\xi\frac{2 u^{0} \delta m(r_{\Omega})}{3 f_{\Omega } r_{\Omega }}~.
\eeq
where $u^{0}=1/\sqrt{1-3x}$. Note that, $\textup{E}^{(0)}_{\textrm{orb}}$ is the orbital energy of the secondary in vacuum Schwarzschild spacetime and $\textup{E}^{(0,1)}_{\textrm{orb}}$ represents the correction to the orbital energy due to presence of the dark matter.
\par
Within this fixed-frequency formalism, we solve \autoref{pp}, \autoref{matpp}, and \autoref{cov_cons_law11} by adopting the Regge–Wheeler formalism. At order $\mathcal{O}(\epsilon)$, the perturbation equations (\autoref{pp}) can be expressed in terms of two master functions, $\Psi_{R}^{(1,0)}$ and $\Psi_{Z}^{(1,0)}$, which satisfy the inhomogeneous Regge–Wheeler and Zerilli equations, respectively \cite{PhysRevD.67.104017}
\beq\label{eq_o_e}
\left[\partial^2_{r_{*}}+\omega^2-V_{R}(r)\right]\Psi_R^{(1,0)} &=S_R^{(1,0)}~\,,\\
\left[\partial^2_{r_{*}}+\omega^2-V_{Z}(r)\right]\Psi_Z^{(1,0)} &=S_Z^{(1,0)}~.
\eeq
Here, $r_{*}$ denotes the tortoise coordinate, defined by $dr_{*}=dr/f$, and $V_{R}$ ($V_{Z}$) is the Regge–Wheeler (Zerilli) potential
\beq
V_{R}(r)&=\frac{f}{r^2}\left[2(n+1)-\frac{6M_{\textup{BH}}}{r}\right]\\
V_{Z}(r)&=\frac{f \left(18 \Mbh^3+18 \Mbh^2 n r+6 \Mbh n^2 r^2+2 (n+1) n^2 r^3\right)}{r^3 (3 \Mbh+n r)^2}
\eeq
where $2(n+1)=l(l+1)$ with $l$ being the angular number. Furthermore, at this order the source terms are purely distributional and can be written as
\beq\label{st10}
S^{(1,0)}_{R,Z}=G^{(1,0)}_{R,Z}\delta_r+F^{(1,0)}_{R,Z}\delta'_r,
\eeq
where $\delta_r\equiv\delta(r-r_\Omega)$ denotes the Dirac delta function, and $G^{(1,0)}_{R,Z}$ and $F^{(1,0)}_{R,Z}$ are functions of the orbital frequency $\Omega$ (see Ref.~\cite{Rahman:2025mip} for their explicit expressions). Note that both the governing equations and the source terms at this order are exactly identical to those in the vacuum Schwarzschild case.
\par
At $\Od$, we need to consider both fluid (\autoref{cov_cons_law11}) and metric (\autoref{matpp}) perturbation. The perturbation equation for fluid perturbation can be described in terms of the master function   $\Psi_{F}^{(1,1)}$, which satisfies a Regge-Wheeler-Zerilli type equation 
\beq\label{eq_o_ez_f}
\left[\frac{d^2}{dr_*^2}+\left(\frac{\omega^2}{c_r^2}-V_{F}(r)\right)\right]\Psi_{F}^{(1,1)}=S^{(1,1)}_{F}\,,
\eeq
with the potential 
\beq
V_{F}(r)&=\frac{\left(c_r^2-1\right){}^2 f'^2}{16 c_r^4}+f(r) \left(\frac{\left(3 c_r^2-1\right) c_t^2 f'}{2 r c_r^4}\right)\\&+f(r) \left(\frac{r^2 \left(c_r^2-1\right) f''(r)+8 (n+1) c_t^2}{4 r^2 c_r^2}\right)\\&+f(r) \left(\frac{f(r) \left(c_t^4-c_r^2 c_t^2\right)}{r^2 c_r^4}\right)~,
\eeq
where $c_{t}$ and $c_{r}$ are the tangential and radial sound speed defined through the relation $\delta p_{t,r}^{lm}=c_{t,r} \delta \rho^{lm}$ where $\delta \rho^{lm}$ is density perturbation and $\delta p_{t,r}^{lm}$ are the tangential (radial) pressure perturbation. In the above derivation, both radial and tangential sound speeds are assumed to be constant. The source term $S^{(1,1)}_{F}$ can be written as  
\beq\label{stD10}
S^{(1,1)}_{F}&=G^{(1,1)}_{F}\delta_r+F^{(1,1)}_{F}\delta'_r+H^{(1,1)}_{F}\delta''_r\\
&+S^{U(1,1)}_{F}\left(\Psi_{Z}^{(1,0)},\partial_r\Psi_{Z}^{(1,0)}\right)~,
\eeq
where the unbounded part of the source $S^{U(1,1)}_{F}$ is a function of Zerilli mater function and its derivative (see \cite{Rahman:2025mip} for the explicit expression of the source term). 
\par
Similar to \autoref{eq_o_e}, the gravitational perturbation at $\Od$ is described in terms of two master function $\Psi_R\tpd$ and $\Psi_Z\tpd$, which satisfies the inhomogeneous Regge-Wheeler and Zerilli equation respectively,
\beq\label{eq_o_ez}
\left[\partial^2_{r_{*}}+\omega^2-V_{R}(r)\right]\Psi_{R}^{(1,1)}&=S^{(1,1)}_{R}\,,
\\
\left[\partial^2_{r_{*}}+\omega^2-V_{Z}(r)\right]\Psi_{Z}^{(1,1)}&=S^{(1,1)}_{Z}\,.
\eeq
with the source term
\beq\label{stD11}
S^{(1,1)}_{R,Z}=G^{(1,1)}_{R,Z}\delta_r+F^{(1,1)}_{R,Z}\delta'_r+H^{(1,1)}_{R,Z}\delta''_r+S^{U(1,1)}_{R,Z}~,
\eeq
Note that, the unbounded part of the source term for Regge-Wheeler equation depends only on the master function 
 $\Psi_R\tpp$ and its derivative
\beq\label{stUR11}
S^{U(1,1)}_{R}=S^{U(1,1)}_{R}\left(\Psi_{R}^{(1,0)},\partial_r\Psi_{R}^{(1,0)}\right)\,.
\eeq
whereas, for the Zerilli equation, the unbounded source term is a function of $\Psi_Z\tpp$ at $\Op$ and $\Psi_F\tpd$ and their derivatives
\beq\label{stUF11}
S^{U(1,1)}_{Z} &=S^{U(1,1)}_{Z,F}\left(\Psi_F\tpd,\partial_r \Psi_F\tpd\right)~,\\
&+S^{U(1,1)}_{Z,G}\left(\Psi_{Z}^{(1,0)},\partial_r\Psi_{Z}^{(1,0)}\right)\,.
\eeq
We solve  \autoref{eq_o_e}, and \autoref{eq_o_ez} with the boundary condition \cite{Martel:2005ir}
\beq\label{bc_grav}
\Psi_{R,Z}^{\pm} =e^{\pm i \omega r_{*}}\,,\quad r_{*}\to\pm \infty\,,
\eeq 
corresponding to ``in'' (``up'') solutions, representing radiative modes ($l>2$) that are purely ingoing ( outgoing) at horizon (infinity).
For the fluid perturbation (\autoref{eq_o_ez_f}), we assume that the influx of radiation and dark matter has negligible effect on the mass and spin of the secondary. This leads to Dirichlet boundary conditions, $\Psi_F\tpd = 0$ at $4\Mbh$ and $r_c$. \par
By solving \autoref{eq_o_e}, \autoref{eq_o_ez_f} and \autoref{eq_o_ez}, we can obtain the expression for the GW flux
 \beq\label{flux_formula}
 \dot{\textup{E}}_{\textrm{GW}}&=\dot{\textup{E}}^{(1,0)}_{\textrm{GW}}+\xi \dot{\textup{E}}^{(1,1)}_{\textrm{GW}}~,\\ 
 &=\left(\dot{\textup{E}}^{(1,0)+}_{\textrm{GW}}+\dot{\textup{E}}^{(1,0)-}_{\textrm{GW}}\right)+\xi \left(\dot{\textup{E}}^{(1,1)+}_{\textrm{GW}}+\dot{\textup{E}}^{(1,1)-}_{\textrm{GW}}\right)~,
 \eeq
 where $\dot{\textup{E}}^{+}_{\textrm{GW}}$ ($\dot{\textup{E}}^{-}_{\textrm{GW}}$) represents the flux at infinity (at the horizon), the expression of which can be written as
\begin{align}
\dot{\textup{E}}^{(1,0)\pm}_{\textrm{GW}} &=\sum_{lm}C_{lm}\left[\omega^2\left|\Psi_Z\tpp\right|^2+4\left|\Psi_R\tpp\right|^2\right]_{r_{*}\to\pm \infty}\,,\label{flux01}\\
\dot{\textup{E}}^{(1,1)\pm}_{\textrm{GW}} &=\delta f^{\pm}_{b}\dot{\textup{E}}^{(1,0)\pm}_{\textrm{GW}}+\sum_{lm}C_{lm}\bigg[\omega^2\textrm{Re}\left(\Psi_Z\tpp \overline{\Psi_Z\tpd}\right)\nonumber\\
&+4\textrm{Re}\left(\Psi_R\tpp \overline{\Psi_R\tpd}\right)\bigg]_{r_{*}\to\pm \infty}\,,\label{flux11}
\end{align}
 where $C_{lm}=(l+2)!/[64\pi (l-2)!]$, $\delta f_b^{+}=\delta f(r_c)/2$, and $\delta f_b^{-}=0$. Note that, within the fixed-frequency formalism, $\omega=m\Omega$ for quasi-circular orbits in \autoref{eq_o_e}, \autoref{eq_o_ez_f}, and \autoref{eq_o_ez}. As a consequence, the Green’s functions associated with the Regge-Wheeler and Zerilli equations are identical at orders $\mathcal{O}(\epsilon)$ and $\mathcal{O}(\e\xi)$. This allows us to extract the modification to the GW flux, $\dot{\textup{E}}^{(1,1)\pm}_{\textrm{GW}}$, induced by the presence of dark matter without resorting to numerical fitting techniques, since the correction can be cleanly isolated within the fixed-frequency formalism. Note that in \autoref{flux11}, part of the modification to the GW flux in the presence of dark matter arises from the redshift effect, $\delta f^{+}_{b}\dot{\textup{E}}^{(1,0)}_{\textrm{GW}}$. The remaining contribution can be attributed to modifications of the background geometry induced by the dark matter mass function and the gravitational radiation sourced by fluid fluctuations. 
\par
In addition to GW emission, the secondary object also loses energy through local interactions with the surrounding dark matter fluid. As the secondary moves through this medium, it generates a local overdensity in its wake. The gravitational interaction between this overdensity and the secondary gives rise to a dissipative force, referred to as \textit{dynamical friction}, through which the secondary loses orbital energy. The corresponding rate of energy loss due to dynamical friction is given by \cite{Barausse:2007ph,Speeney:2022ryg}
\begin{equation}\label{dynamic_friction}
\dot{\textup{E}}_{\textrm{DF}} \approx 3\frac{4\pi}{\sqrt{x}}\frac{1+x}{1-x}m_p^2 \rho_{\textrm{BH}}\,.
\end{equation}
\par
As a result of the GW emission and dynamic friction, the secondary object loses orbital energy and the object slowly inspirals towards the primary. The corresponding flux-balance law can be written as follows
\begin{equation}\label{flux_balance_law_1}
\dot{\textup{E}}_{\textrm{orb}} = -\dot{\textup{E}}^{(0,1)}_{\textrm{GW}} - \xi\left[\dot{\textup{E}}^{(1,1)}_{\textrm{GW}}+\dot{\textup{E}}_{\textrm{DF}}\right].
\end{equation}
Note that, $\dot{\textup{E}}^{(0,1)}_{\textrm{GW}}$ represents the energy flux in vacuum Schwarzschild spacetime. The coefficient of the dark matter parameter $\xi$ in the above equation thus describes the modification of the flux balance law in the presence of dark matter. As a result of the orbital energy loss, the orbital frequency of the object slowly evolves.  The corresponding evolution equation of the secondary can be written in terms of action-angle variables as follows
 \beq\label{frequency}
\dot\phi_p &=\frac{d\phi_p}{dt}\equiv \Omega~,\\
\dot\Omega &=\frac{d\Omega}{dt}=\e \F^{(1,0)}(\Omega)+\e\xi \F^{(1,1)}(\Omega)~,
 \eeq
where the orbital phase $\phi_p$ is the orbital phase evolves in fast orbital time scale $T_o\sim \Mbh$, and the frequency changes in slow inspiral time scale $T_i\sim \Mbh/\e$. Substituting the above equation in \autoref{flux_balance_law_1}, and using the relation $\dot{\textup{E}}_{\textrm{orb}}=\Omega \partial_{\phi_p}\textup{E}_{\textrm{orb}}+\dot\Omega \partial_{\Omega} \textup{E}_{\textrm{orb}}$, we can obtain the expression for the quantities $\F^{(1,0)}(\Omega)$ and $\F^{(1,1)}(\Omega)$ as 
\beq\label{flux_bal_adia}
\F^{(1,0)}&=-\frac{\dot{\textup{E}}_{\textrm{GW}}\tpp}{\partial_\Omega \textup{E}^{(0)}_{\textrm{orb}}}\,,\\
\F^{(1,1)}&=-\frac{\dot{\textup{E}}_{\textrm{GW}}\tpd+\F^{(1,1)} \partial_\Omega \textup{E}^{(0,1)}_{\textrm{orb}}}{\partial_\Omega \textup{E}^{(0)}_{\textrm{orb}}}-\frac{\dot{\textup{E}}_{\textrm{DF}}\tpd}{\partial_\Omega \textup{E}^{(0)}_{\textrm{orb}}}~.
\eeq
Note that the term $\F^{(1,1)}$ represents the modification of the evolution equation due to the presence of dark matter. By solving the equations in \autoref{frequency} simultaneously, we can obtain the evolution of EMRIs in the dark matter environment.
 \section{EMRI in scalar Gauss-Bonnet gravity } \label{secIII}
sGB gravity is a subclass of scalar–tensor theories in which a (massless) scalar field couples non-minimally to the Gauss–Bonnet invariant. The action can be written as \cite{Maselli:2020zgv, Spiers:2023cva}
\begin{align}\label{scalaraction1}
S[\mathbf{g}_{\mu\nu},\varphi, \mathbf{\Psi}] = S_{0}[\mathbf{g}_{\mu\nu},\varphi]+\alpha S_{c}[\mathbf{g}_{\mu\nu},\varphi]+S_{\textup{mat}}[\mathbf{g}_{\mu\nu},\varphi, \mathbf{\Psi}],
\end{align}
where
\begin{align}
S_0[\mathbf{g}_{\mu\nu},\varphi] = \int d^{4}x\frac{\sqrt{-\mathbf{g}}}{16\pi}\Big(R-\frac{1}{2}\partial_{\mu}\varphi \partial^{\mu}\varphi\Big),
\end{align}
where $R$ is the Ricci scalar, $\varphi$ is a real scalar field, the parameter $\alpha S_{c}$ parametrizes the non-minimal interaction between the scalar field and the spacetime metric $\mathbf{g}_{\mu\nu}$ that enables us to write the action (\autoref{scalaraction1}) alternatively in the following fashion:
\begin{align}
S = S_0 + \frac{\alpha}{4}\int \frac{\sqrt{-\mathbf{g}}}{16\pi}\kappa (\varphi)\mathcal{G}+S_{\textup{mat}}.
\end{align}
$\alpha$ (with dimension of $[\textup{mass}]^{n}$, $n\geq 1$) is the Gauss–Bonnet coupling constant, $\kappa(\varphi)$ is a dimensionless coupling function, $\mathcal{G} = R^{2}-4R_{\mu\nu}R^{\mu\nu}+R_{\mu\nu\alpha\beta}R^{\mu\nu\alpha\beta}$ is the Gauss–Bonnet invariant term and $S_{\textup{mat}}$ is the action for matter fields $\mathbf{\Psi}$. 

If one varies the action (\autoref{scalaraction1}) with respect to metric, one can obtain field equations \cite{Maselli:2020zgv, Spiers:2023cva}:
\begin{align}\label{fieldequ2}
G_{\mu\nu}[\mathbf{g}_{\alpha\beta}] = \textup{T}^{\textup{scal}}_{\mu\nu}[\varphi]+\alpha \textup{T}^{\textup{c}}_{\mu\nu}[\mathbf{g}_{\alpha\beta},\varphi] + \textup{T}^{\textup{mat}}[\mathbf{g}_{\alpha\beta},\varphi,\mathbf{\Psi}],
\end{align}
where $G_{\mu\nu}[\mathbf{g}_{\alpha\beta}]$ is defined as the Einstein tensor and
\begin{align}
\textup{T}^{j}_{\mu\nu} = -\frac{8\pi}{\sqrt{-\mathbf{g}}}\frac{\delta S_{(j)}}{\delta g^{\mu\nu}}; \textup{T}^{\textup{scal}}_{\mu\nu}[\varphi]=\frac{1}{2}\partial_{\mu}\varphi\partial_{\nu}\varphi - \frac{1}{4}g_{\mu\nu}\partial_{\alpha}\varphi\partial^{\alpha}\varphi~,
\end{align}
with $j\in (\textup{c}, \textup{mat})$. Now for the scalar field equation, one can vary the action (\autoref{scalaraction1}) with respect to the field $\varphi$ \cite{Spiers:2023cva}:
\begin{align}\label{scalecoup}
\square\varphi = \alpha\mathsf{\Sigma}_{\textup{c}}+\mathsf{\Sigma},
\end{align}
where $\square$ is the D’Alambertian operator for the Schwarzschild metric ($\Box \equiv \nabla_\mu \nabla^\mu$), and $\mathbf{g}$ is the metric determinant, and
\begin{align}
\mathsf{\Sigma} = -\frac{16\pi}{\sqrt{-\mathbf{g}}}\frac{\delta S_{\textup{mat}}}{\delta\varphi} \hspace{0.2cm} ; \hspace{0.2cm} \mathsf{\Sigma}_{\textup{c}} = -\frac{16\pi}{\sqrt{-\mathbf{g}}} \frac{\delta S_{\textup{c}}}{\delta\varphi}.
\end{align}
Further, it has been noted that black holes in this theory can evade no-hair theorem and can allow a scalar field to exist outside its event horizon \cite{PhysRevLett.112.251102}. However, the charge of the scalar field itself depends on the mass of the black hole. The larger the black hole mass is, the less charge it carries.
 \par
If we consider an EMRI in such theories, the primary object  is a supermassive black hole, whose spacetime curvature scales as $M_{\textup{BH}}^{-2}$. Since the scalar field in sGB gravity is sourced by curvature, massive black holes carry smaller scalar charge, consequently, the sGB-induced corrections to the background geometry are suppressed by powers of $\alpha/M_{\textup{BH}}^{2}$ (for $n=2$) \cite{Maselli:2020zgv}. In other words, deviations from the Schwarzschild geometry are governed by the dimensionless parameter $\zeta \equiv \frac{\alpha}{M_{\textup{BH}}^{n}} = q^{n}\frac{\alpha}{m_{p}^{n}}$ where $m_p$ is the mass of the secondary object and $q \ll 1$ is the mass-ratio.
 Requiring $\alpha/m_{p}^{n} < 1$ in order to remain consistent with existing astrophysical constraints implies $\zeta \ll 1$, thereby justifying the use of the Schwarzschild metric to describe the exterior spacetime \cite{Maselli:2021men}. Therefore, for supermassive primaries in EMRI systems, this dimensionless coupling is extremely small, and consequently the background spacetime can be accurately described by the vacuum Schwarzschild metric, i.e., $g_{\mu\nu} = \bar{g}_{\mu\nu}$. This also implies that unlike the dark matter case, there is no need to introduce a background-level modification of the metric analogous to $h_{\mu\nu}^{(0,1)}$ in \autoref{DM_spacetime}.
 \par
Moreover, the sGB effects enter through the dynamics of the scalar field and its coupling to the motion of the secondary. We consider an EMRI with $m_{p}\ll M_{BH}$, modelling the secondary using the skeletonized (point-particle) approximation that replaces the matter-action $S_{\textup{mat}}$ by the point-particle action $S_{\textup{p}}$ \cite{Maselli:2020zgv, Zi:2025lio}:
 \begin{align}\label{ppact}
S_{\textup{p}} = \int -m(\varphi)\sqrt{g_{\alpha\beta}\frac{dz^{\alpha}}{d\tau}\frac{dz^{\beta}}{d\tau}} d\tau,
 \end{align}
where $z^\alpha(\tau)$ denotes the particle trajectory, with a scalar-field--dependent mass $m(\varphi)$ and $\tau$ is the proper time. The mass function $m(\varphi)$ is determined by evaluating the scalar field at the particle’s position, as sourced by the dynamics of the moving secondary. 

Given the point particle action (\autoref{ppact}), if we vary this with respect to metric $\mathbf{g}_{\mu\nu}$, we obtain
\begin{align}
\textup{T}^{\mu\nu}_{\textup{p}} =   \int d\tau m (\varphi)\frac{\delta^{(4)}(x-z(\tau))}{\sqrt{-\mathbf{g}}} {u}_{\mu} {u}_{\nu}.
\end{align}
This is stress-energy tensor of scalar charged point particle. If we vary action with respect to scalar field $\varphi$, it gives
\begin{align}
\mathsf{\Sigma} =16\pi\int m'(\varphi)\frac{\delta^{4}(x-x(\tau))}{\sqrt{-\mathbf{g}}}d\tau.
\end{align}
This provides us with the scalar charge density of the point particle. Apart from this, The scalar stress--energy tensor $\textup{T}^{\mathrm{scal}}_{\mu\nu}$ is second order in perturbations about $\varphi=\varphi_0$ and can therefore be consistently neglected.

\subsection{EMRI systems in sGB gravity}
As argued above, the smallness of the dimensionless coupling in EMRI systems allows the background geometry to be well described by the Schwarzschild spacetime. The metric perturbations, therefore, enter only at leading-order in the mass ratio. In particular, the absence of any background-level deformation implies that
$h_{\mu\nu}^{(0,1)}$ is zero and, consequently, no mixed-order contributions such as $h_{\mu\nu}^{(1,1)}$ appear. As a result, the metric can be expanded as
\begin{align}\label{metexp}
\mathbf{g}_{\mu\nu} = \bar{g}_{\mu\nu}+\epsilon h^{(1,0)}_{\mu\nu},
\end{align}
with 
$\epsilon$ denoting the order in the mass-ratio expansion, $\bar{g}_{\mu\nu}$ is the Schwarzschild metric and $h^{(1,0)}_{\mu\nu}$ is the first-order perturbation to the background. Similarly, one can also write down the scalar field expansion
\begin{align}\label{scalexp}
\varphi = \bar\varphi + \epsilon \varphi^{(1,0)},
\end{align}
where $\bar\varphi$ denotes the scalar contribution from the isolated supermassive black hole action $S_0$, and can be set to zero without loss of generality. The term $\varphi^{(1,0)}$ corresponds to the first-order perturbation to the scalar field. Furthermore, since the point particle stress-energy tensor of the inspiralling object carries mass function $m(\varphi)$, we expand 
\begin{align}\label{massfunscal}
m(\varphi) = \bar{m}+m_{(1,0)}\varphi,
\end{align}
Here ($\bar{m}, m_{(1,0)}$) are constant coefficients with $\bar{m}=m_{p}$, i.e., mass of the secondary object in GR with $\varphi=0$. If we put  (\autoref{scalexp}) into  (\autoref{massfunscal}), we obtain the mass function $m(\varphi)$ to the linear order in $\epsilon$:
\begin{align}\label{massfun2}
m(\varphi) = m_{p}+\epsilon \textup{m}_{(1,0)}\varphi_{(1,0)}.
\end{align}
Note that in modified gravity theories like scalar-tensor, $m_{p}$ does not reflect the total mass of the secondary object, instead the correction can arise through $\textup{m}_{(1,0)}$.

Let us now expand  (\autoref{fieldequ2}) to the linear-order in mass-ratio that gives rise the same expression equivalent to  (\autoref{pp}), i.e., same as in the GR case, and written as \cite{Maselli:2021men,Maselli:2020zgv,Barsanti:2022ana}
\begin{align}\label{pp1}
\delta G_{\mu\nu}^{(1,0)} = 8\pi \T^{p(1,0)}_{\mu\nu}=8\pi m_{p}\int\frac{\delta^{4}(x-z(\tau))}{\sqrt{-\mathbf{g}}}u^{\mu}u^\nu d\tau~.
\end{align}
Further, we now also understand that under the assumption coupling being very small, the scalar field obeys a Klein--Gordon equation sourced solely by the motion of the secondary compact object. The resultant expression is given by
\begin{align}\label{KGEq1}
\square\varphi^{(1,0)} = 16\pi\int \textup{m}_{(1,0)}\frac{\delta^{4}(x-z(\tau))}{\sqrt{-\mathbf{g}}}d\tau,
\end{align}
where $\square = \nabla_{\mu}\nabla^{\mu}$ is covariant derivative associated with the Schwarzschild metric. At $\mathcal{O}(\epsilon)$, the correction $\textup{m}_{(1,0)}$ encodes the scalar charge of the secondary, as shown in \cite{Maselli:2020zgv, Maselli:2021men, Barsanti:2022ana}, through a buffer-region expansion $\textup{m}_{p} \ll \hat r \ll M$. In other words, for the EMRI system under consideration, the secondary of mass $m_{\mathrm{p}} \ll M_{\mathrm{BH}}$ moves in the spacetime of the primary while carrying a massless scalar field, admitting a local expansion in a frame centered on the secondary, which is written as
\begin{align}\label{fieldexp}
\varphi^{(1,0)} = \frac{m_{p}q_s}{\tilde r} + \mathcal{O}\Big(\frac{m^{2}_{p}}{\tilde{r}^{2}} \Big),
\end{align}
where $\tilde{r}$ is the local radial coordinate and $q_s$ is the dimensionless scalar charge. Consequently, using Eqs. (\autoref{KGEq1}) and (\autoref{fieldexp}), in the buffer zone, the matching solution gives rise $\textup{m}_{(1,0)}$:
\begin{align}
\textup{m}_{(1,0)} = -\frac{m_{p}q_s}{4}.
\end{align}
which reduces the  (\autoref{KGEq1}) in the following form \cite{Maselli:2020zgv, Zi:2025lio}:
\begin{equation} \label{scalareq}
\square \varphi^{(1,0)} = -4\pi q_s m_{p} \int \frac{\delta^{(4)}\!\left(x - z(\tau)\right)}{\sqrt{-\mathbf{g}}} d\tau.
\end{equation}
 (\autoref{scalareq}) includes a source term in the scalar field equation, which is governed by the scalar charge of the secondary and further controls the resulting deviations from GR in the EMRI evolution. For the related details and assumptions highlighted here, readers are also encouraged to read \cite{Maselli:2020zgv, Spiers:2023cva, Zi:2025lio, Maselli:2021men, Sotiriou:2013qea, Julie:2019sab}. \par
Due to the presence of scalar field and gravitational self-force effects, the secondary object follows a forced geodesic orbit in terms of the background metric \cite{Spiers:2023cva}
\beq\label{sf_eqn_sGB}
u^{\mu}_{;\nu}u^{\nu}&=
\e F_{G(1,0)}^{\mu}+\e q_s F_{S(1,1)}^{\mu}~,
\eeq
where $F_{G(1,0)}^{\mu}$ is presented in \autoref{sf_eqn} and $F_{S(1,0)}^{\mu}$ is given by 
\beq\label{sf_eqn_sGB}
 F_{S(1,1)}^{\mu}=-\frac{1}{4} P^{\mu\nu} \nabla_{\nu}\varphi^{R(1,0)}~.
\eeq
Here, $\varphi^{R(1,0)}$ is obtained by removing certain singular pieces from the physical scalar field $\varphi$. Due to the dissipative pieces of the force term $F_{G(1,0)}^{\mu}$ and $F_{S(1,1)}^{\mu}$, the object loses orbital energy and angular momentum  and the object enters the inspiral phase. As in the case of dark matter, in this paper, we focus on the leading order modification to the orbital evolution of EMRIs in the presence of the scalar field.

\subsection{Orbital evolution in sGB gravity}
Similar to the dark matter scenario, we adopt the \textit{fixed-frequency formalism} \cite{Mathews:2021rod} to describe the perturbed orbit
\beq\label{r_expand_sGB}
\Omega = \Omega_0\,,\quad
r = r_\Omega +\epsilon r_{(1,0)}(\Omega) + \epsilon q_s  r^{S}_{(1,1)}(\Omega),\quad
\eeq
where $r_{(1,0)}(\Omega)$ is given in \autoref{r_expand} and 
\beq
\quad r^S_{(1,1)}=-\frac{F^{r}_{S(1,1)}(1-3x)}{3\Omega^2 f_{\Omega}}~.
\eeq
In this fixed-frequency formalism, we solve the perturbation equation \autoref{pp1} and \autoref{KGEq1} through Regge-Wheeler formalism. Note that, the gravitational perturbation equation \autoref{pp1} is same as \autoref{pp}. Thus, the gravitational perturbation equation can be described in terms of two master function $\Psi^{(1,0)}_R$ and  $\Psi^{(1,0)}_Z$, which satisfies \autoref{eq_o_e}. The perturbation of the scalar field can also be described in terms of a master function $\Psi^{(1,0)}_S$ whicg satisfies a Regge-Wheeler type equation \cite{Sago:2002fe, Maselli:2020zgv}
\begin{align}\label{RWZpert1}
\left[\partial^2_{r_{*}}+\omega^2-V_{S}(r)\right]\Psi_S^{(1,0)} &=S_S^{(1,0)}~,
\end{align}
where $V_{S}(r)$ is the given by
\begin{align}\label{potmatirx}
V_{S}(r)= f(r) \Big(\frac{2(n+1)}{r^{2}}+\frac{2M_{\textup{BH}}}{r^{3}}\Big),
\end{align}
and the  source term is given by
\begin{align}\label{sourcescal}
S_S^{(1,0)} = -q_{s}\textup{m}_{p}f(r_\Omega)\sqrt{r_\Omega-3M_{\textup{BH}}}\frac{4\pi P_{\ell m}(\pi/2)}{r_\Omega^{3/2}}\delta_r,
\end{align}
with $P_{\ell m}(\theta)$ being the Lengendre Polynomial. We solve the above equation with the boundary condition that $\Psi_S^{(1,0)}$ is purely incoming at the event horizon and is purely outgoing at the infinity.

Finally, one can compute energy fluxes at the horizon and infinity. The expression for the GW flux is given in \autoref{flux01}, whereas the scalar wave flux  is given by
\beq\label{fluxesGB}
q_s^2\dot{\textup{E}}^{(1,0)}_{\textrm{SW}}&=q_s^2\dot{\textup{E}}^{(1,0)+}_{SW}+q_s^2\dot{\textup{E}}^{(1,0)-}_{\textrm{SW}}~, \\
q_s^2\dot{\textup{E}}^{(1,0)\pm}_{\textrm{SW}}&= \frac{1}{32\pi}\sum_{lm}\left[\omega^2 \left|\Psi_S\tpp\right|^2\right]_{r_{*}\to\pm \infty}~.
\eeq
As discussed above, $\dot{\textup{E}}_{\textup{GW}}\tpp$ provides the energy fluxes for generated by an EMRI in GR  and $\dot{\textup{E}}_{\textup{SW}}$ gives corrections to the energy fluxes because of the scalar radiation sourced by the scalar charge. \par
Due to the presence of both scalar and GW radiation, the secondary object loses orbital energy
In that case, the secondary satisfies the following energy balance law \cite{Speeney:2022ryg}
\beq\label{flux_balance_law_sGB}
\dot{\textup{E}}_{\textrm{orb}} = -\dot{\textup{E}}_{\textrm{GW}}\tpp - q_s^2\dot{\textup{E}}_{\textrm{SW}}\tpp~.
\eeq
As a result of the orbital energy loss, the orbital frequency of the object slowly evolves.  The corresponding evolution equation of the secondary can be written as
 \beq\label{frequency_1}
\dot\phi_p &=\frac{d\phi_p}{dt}\equiv \Omega~,\\
\dot\Omega &=\frac{d\Omega}{dt}=\e \F^{(1,0)}(\Omega)+\e q_s^2 \F^{S(1,1)}(\Omega)~,
 \eeq
where the expression for the $\F^{(1,0)}(\Omega)$ and $\F^{S(1,1)} (\Omega)$ can be obtained by substituting the above equation in \autoref{flux_balance_law_sGB}, and using the relation $\dot{\textup{E}}_{\textrm{orb}}=\Omega \partial_{\phi_p}\textup{E}_{\textrm{orb}}+\dot\Omega \partial_{\Omega} \textup{E}_{\textrm{orb}}$ as in the dark matter case., which can be written as 
\beq\label{flux_bal_adia}
\F^{(1,0)}=-\frac{\dot{\textup{E}}_{\textrm{GW}}\tpp}{\partial_\Omega \textup{E}^{(0)}_{\textrm{orb}}}\,,\quad
\F^{S(1,1)}=-\frac{\dot{\textup{E}}_{\textrm{SW}}\tpp}{\partial_\Omega \textup{E}^{(0)}_{\textrm{orb}}}~.
\eeq
Similar to dark matter scenario, we solve \autoref{frequency_1} to find the evolution of the EMRI in sGB theory. In the presence of the scalar radiation due to the scalar charge carried by the secondary object speeds up the inspiral process as compared to the vacuum GR case.
\par
The stringent constraints on such scalar charge from future LISA observations can therefore be translated into bounds on beyond GR parameters, while simultaneously enabling a more physical interpretation of beyond-vacuum GR effects and their disentanglement from astrophysical environmental influences.
\section{Waveform and constraint on scalar charge beyond-vacuum}\label{secIV}
In this section, we introduce the formula of quadrupole waveform to compute the constraint on scalar charge in beyond-vacuum setting.

Once the inspiraling trajectories of secodary are obtained from \autoref{flux_balance_law_1} or \autoref{flux_balance_law_sGB}, two polarizations can then be expressed in terms of the derivatives of quadrupole moment of source as
\begin{eqnarray}\label{amplitude}
h^+ &=& -\frac{1}{2}\left(\ddot{I}_{11} - \ddot{I}_{22}\right) (1 + \cos^2\iota)
        = \mathcal{A} (1 + \cos^2\iota) \cos[2\Phi(t)], \nonumber \\
h^\times &=& 2\ddot{I}_{12} = -2\mathcal{A} \cos\iota\, \sin[2\Phi(t)],
\end{eqnarray}
where \( \mathcal{A} = (\Mbh\omega)^{2/3} m_p / d_L \) with secondary object mass $m_p$, and \( \iota \) is the angle of the orbital plane relative to the line of sight,  $\Phi$ is the GW phase. The source position and 
orientation are further specified by the four angles \( (\theta_S, \phi_S, \theta_L, \phi_L) \), following the convention of Ref.~\cite{Barack:2003fp}.
The second time derivative of the mass quadrupole moment is determined by the source 
stress-energy tensor~\cite{Babak:2006uv},
\begin{equation}\label{eq:quad:stress-tensor}
I_{ij} = \int d^3x\, T^{tt}(t, x^i)\, x^i x^j = m_p\, y^i(t)\, y^j(t),
\end{equation}
where \( T^{tt}(t, x^i) = m_p \delta^{(3)}(x^i - z^i(t)) \), 
\( x^i \) are Cartesian spatial coordinates, and \( y^i(t) \) 
denotes the worldline of the secondary compact object.
Under low-frequency approximation, the amplitudes of GW responded by LISA can be written as  
\begin{equation}\label{antenna}
h_{\textup{I,II}} = \sum_n \frac{\sqrt{3}}{2} \Big[F^+_{\textup{I,II}} (t)A^+_n(t) + F^\times_{\textup{I,II}} (t)A^\times_n(t) \Big]\;,
\end{equation}
where the quantities $F^+_{\textup{I,II}}$ are the antenna pattern functions, their detailed expressions can be referred to Ref.~\cite{Cutler:1997ta}.
\begin{figure*}[t!]
	\centering
	\minipage{0.48\textwidth}
	\includegraphics[width=\linewidth]{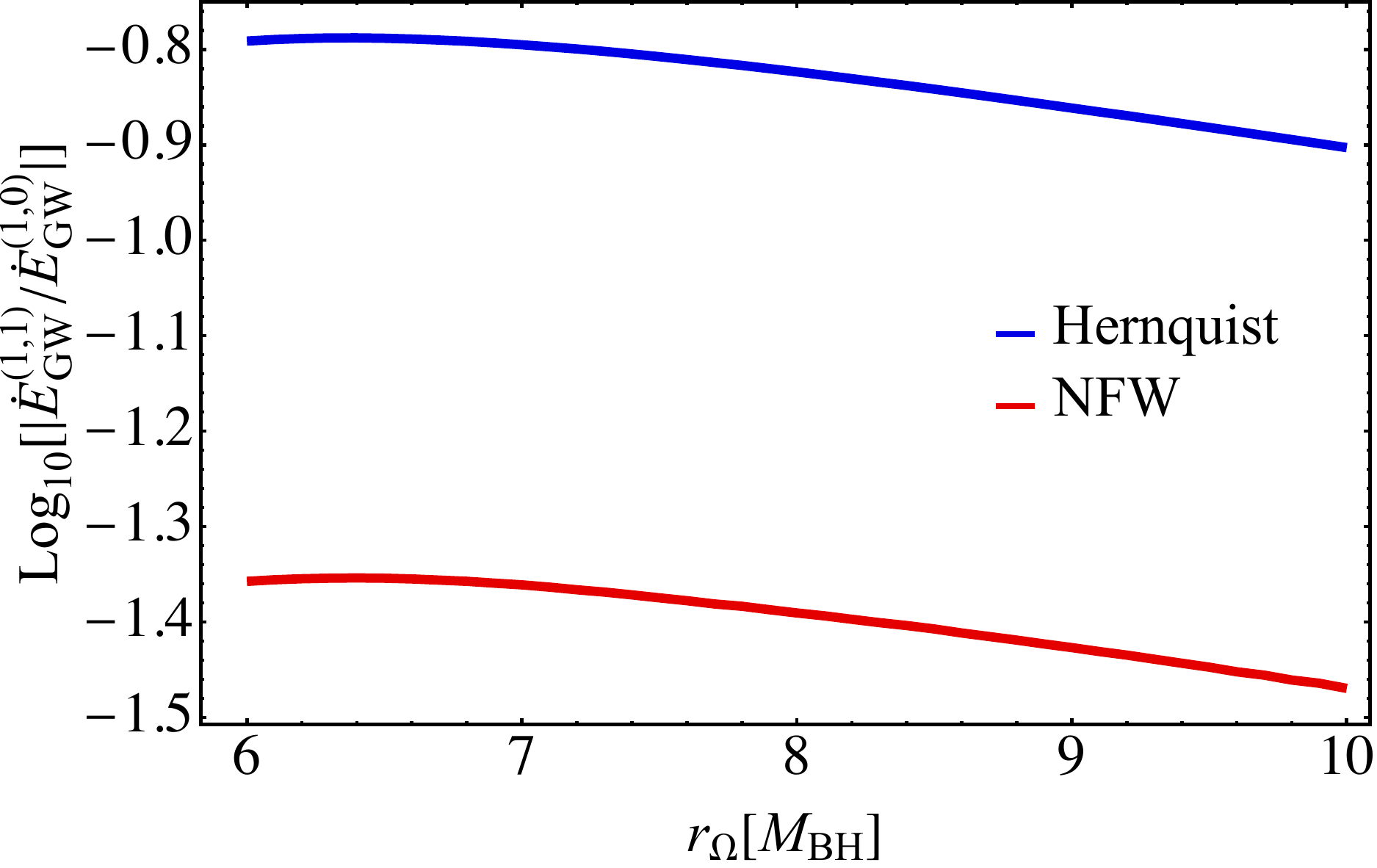}
	\endminipage\hfill
    \minipage{0.48\textwidth}
	\includegraphics[width=\linewidth]{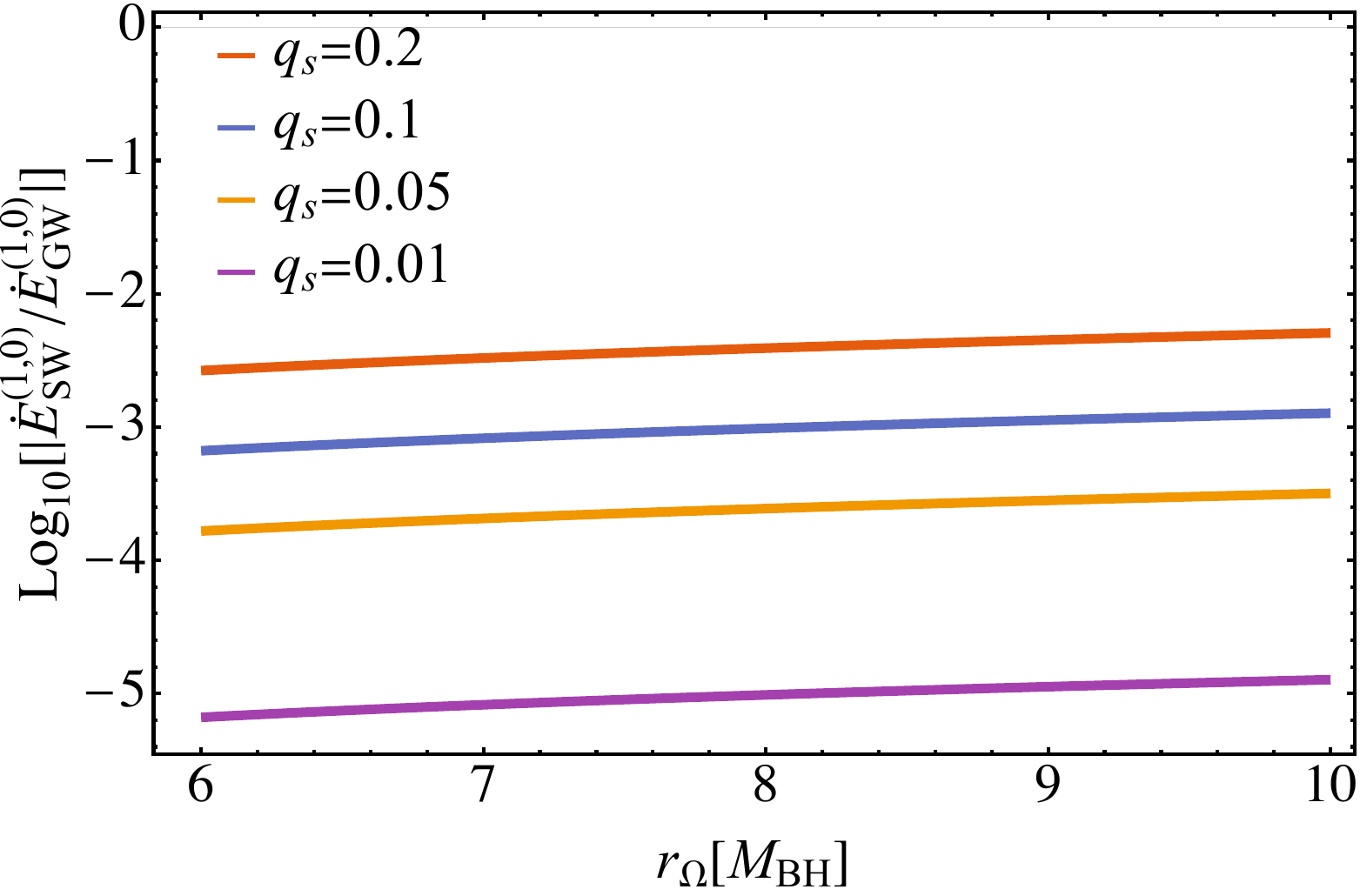}
	\endminipage
	\caption{The energy fluxes as functions of the orbital radius \(r_{\Omega}/\Mbh\) is presented. Left panel: The absolute value of the ratio between the dark-matter-induced correction to the gravitational wave flux, $\dot{\textup{E}}_{\textrm{GW}}^{(1,1)}$, and the vacuum gravitational wave flux, $\dot{\textup{E}}_{\textrm{GW}}^{(1,0)}$, as a function of $r_\Omega/\Mbh$ for the Hernquist and NFW profiles is presented. The numerical parameters used in this flux calculation are the same as those listed in \autoref{tab:Fitting_Parameters}. Right panel: the absolute value of the ratio between the scalar radiation flux, $\dot{\textup{E}}_{\textrm{SW}}^{(1,0)}$, and the gravitational wave flux, $\dot{\textup{E}}_{\textrm{GW}}^{(1,0)}$, in sGB theory for different values of the scalar charge, shown as a function of $r_\Omega/\Mbh$.
 }\label{fig:flux}
\end{figure*}
Finally, we estimate EMRI parameters to constrain the scalar charge and dark matter environment using the Fisher information matrix (FIM). In GW inference, 
the FIM provides leading-order estimates of statistical uncertainties under 
the assumptions of stationary, Gaussian noise, high signal-to-noise ratio (SNR), 
and weak or uniform priors~\cite{Vallisneri:2007ev}. In this regime, the 
posterior distribution is well approximated by a multivariate normal centered 
on the true parameters, with a covariance matrix given by the inverse of the FIM.
We consider the parameter vector
\begin{equation}\label{eq:parameters:vector}
\boldsymbol{\theta} = 
\{ \Mbh, m_p, r_p, \mfa, \mfb, \mfc, a_0, \Mh, q_s, \Phi_0, \theta_S, \phi_S, \theta_K, \phi_K, d_L \} \;,
\end{equation}
where $r_p=r_{\Omega}$,  $(\mfa, \mfb, \mfc)$ are fitting Parameters for the dark matter spike profile of the Hernquist and NFW cases in \autoref{tab:Fitting_Parameters}, $a_0$ is the length scale of the the dark matter halo, $\Mh$ is the mass of the dark matter halo, $q_s$ is the scalar charge related to the family of scalar-tensor gravity theory, and $d_L$ is the luminosity distance. The  initial phase have the following relationship between point-particle's orbital phase and GW phase $\Phi_0=2\phi_{p,0}$.

The Fisher information matrix $\Gamma_{ij}$ can be expressed as
\begin{equation}
\Gamma_{ij} \equiv 
\left( \frac{\partial h}{\partial \theta_i} \,\bigg|\, 
       \frac{\partial h}{\partial \theta_j} \right),
\qquad i,j = 1, \dots, 15 \, .
\end{equation}
where $(\cdot|\cdot)$ denotes the usual inner product weighted by the detector noise spectral density
\begin{align}\label{ovp2}
(h_{1}|h_{2}) = 4 \textrm{Re}\left[ \Large\int_{0}^{\infty} \frac{\tilde{h}_{1}(f) \tilde{h}^{*}_{2}(f)}{S_{n}(f)}\right] ,
\end{align} 
with $\tilde h_{i}(f)$ being the Fourier transformation of $h_{i}(t)$ and $S_{n}(f)$ is the one-sided noise spectral density.
The covariance matrix is then given by
\begin{equation}\label{eq:cov:matrix}
\Sigma_{ij} \equiv 
\langle \delta\theta_i \, \delta\theta_j \rangle
= \left( \Gamma^{-1} \right)_{ij},
\end{equation}
and the corresponding $1\sigma$ uncertainties on each parameter is obtained from its diagonal elements,
\begin{equation}\label{sigma:fim}
\sigma_{\theta_i} = \sqrt{\Sigma_{ii}} \, .
\end{equation}
This framework provides a rough estimate of LISA's sensitivity to the scalar charge $q_s$. 
Moreover, the off-diagonal components of $\Sigma_{ij}$ quantify correlations 
among intrinsic and extrinsic parameters, offering insight into possible 
degeneracies in the EMRIs parameter space.

Additionally, we also define dephasing and mismatch to assess the effect of dark matter on EMRIs orbital dynamics.
First, the dephasing can be given by the integration of difference of two orbital frequencies under the influence of dark matter
\begin{eqnarray}
\Delta \Phi^{\rm GR, sGB} & = & 2\int_0^{t} \Big(\Omega^{\rm GR} \big[t\big] - \Omega^{\rm sGB} \big[t\big] \Big) dt\;,\\
\Delta \Phi^{\rm DM, sGB} &=& 2\int_0^{t} \Big(\Omega^{\rm DM} \big[t\big] - \Omega^{\rm sGB} \big[t\big] \Big) dt\;,\\
\Delta \Phi^{\rm DM, vac} &=& 2\int_0^{t} \Big(\Omega^{\rm DM} \big[t\big] - \Omega^{\rm vac} \big[t\big] \Big) dt\;,
\end{eqnarray}
where quantity $\Delta \Phi^{\rm GR, sGB}$ is the GW phase difference between the GR and sGB theories,  $\Delta \Phi^{\rm DM, sGB}$ is the phase deviation for EMRI signals from the dark matter environment and sGB theory and $\Delta \Phi^{\rm DM, vac}$ represents the GW phase difference of EMRIs in the vacuum and dark matter environment. The phase of GW can be obtained with orbital frequency affected by different environments and theories, the quantities $\Omega[t]$ with superscripts ``${\rm GR, DM, sGB}$'' are obtained by EMRIs fluxes in the standard GR, dark matter environments and sGB theory.

As already emphasized in the literature, phase dephasing alone is a useful indicator of the potential detectability of specific physical effects, but it does not by itself provide a robust, detector–dependent measure of observability for LISA-like experiments. To quantify how strongly a given EMRI signal is recorded by the detector, and to assess whether two signals can be distinguished, we instead use the standard noise-weighted mismatch as a figure of merit. For two waveforms $h_a$ and $h_b$, the mismatch is defined as
\begin{align}\label{overlap}
\mathcal{M} \equiv 1-\mathcal{O}(h_a, h_b)
= 1- \frac{(h_a \vert h_b)}{\sqrt{(h_a \vert h_a)(h_b \vert h_b)}} \;.
\end{align}
 The SNR of EMRI signal in a GW detector is then
\begin{equation}
\mathrm{SNR} \equiv \rho = \sqrt{(h_a|h_a)} .
\end{equation}
A typical EMRI signal is considered detectable when its SNR satisfies $\rho \gtrsim 20$~\cite{Babak:2017tow,Fan:2020zhy}, while two competing templates can be distinguished if their mismatch obeys the rule-of-thumb criterion $\mathcal{M} \gtrsim 1/(2\rho^2)$~\cite{Flanagan:1997kp,Lindblom:2008cm}.

In this work, the gravitational wave fluxes and inspiral dynamics are computed within a fully relativistic framework, whereas the GW strain is constructed using a quadrupole formula. This hybrid treatment does not qualitatively affect the dephasing analysis; employing a fully relativistic waveform model would primarily induce modest quantitative shifts in the inferred mismatches and Fisher-matrix forecasts~\cite{Katz:2021yft,Mitra:2025tag}.

\section{Result}\label{result}
In this section, we present the energy fluxes, including the modifications induced by dark matter, the corresponding EMRI waveforms, the correlations between the dark matter effects and the waveform features, and the resulting constraints on the scalar charge beyond the vacuum case. Throughout this analysis, we fix the source-direction and angular-momentum orientation of the EMRIs to $\theta_S=\pi/3$, $\phi_S=\pi/4$, $\theta_K=\pi/3$, and $\phi_K=\pi/4$. The test particle orbital radius $r_p=r_{\Omega}$ is chosen such that the circular EMRIs undergo one year of adiabatic evolution before the final plunge. We consider an EMRI with component masses $\Mbh=10^6 M_\odot$ and $m_p=10 M_\odot$ embedded in two representative dark matter halo models, namely the NFW and Hernquist profiles. The luminosity distance $d_L$ is treated as a free parameter and adjusted to set the expected SNR in the parameter-estimation analysis.


\begin{figure*}[htb!]
\centering
\includegraphics[width=3.2in, height=2.2in]{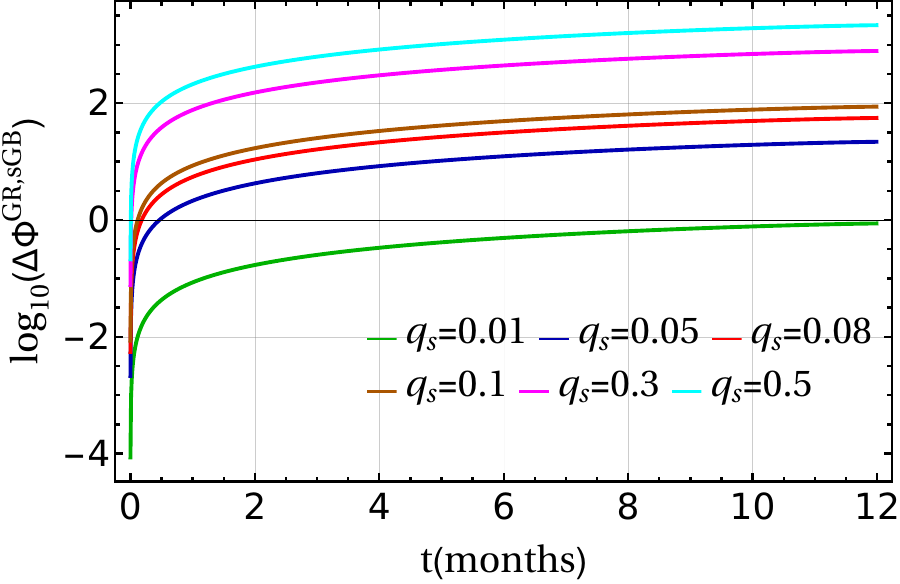}
\includegraphics[width=3.2in, height=2.3in]{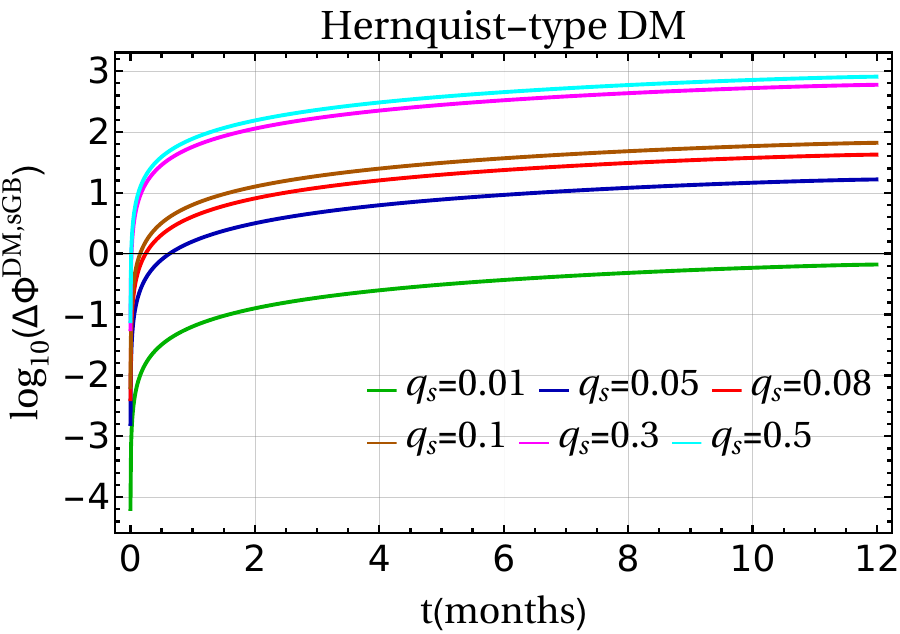}
\includegraphics[width=3.2in, height=2.2in]{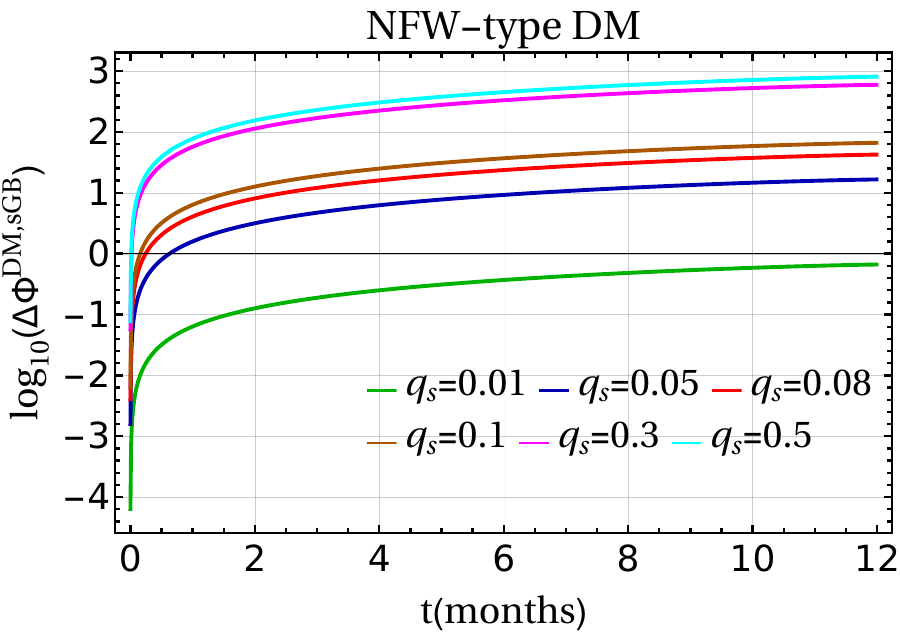}
\includegraphics[width=3.2in, height=2.1in]{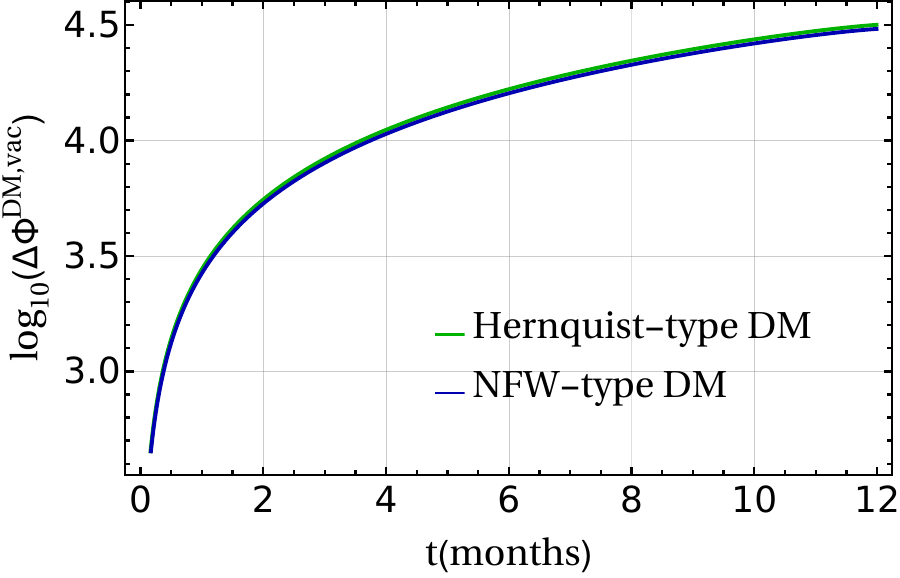}
\caption{The logarithm of dephasing for different environmental configurations as a function of observation time is plotted, which considers four comparison cases: the sGB theory and the standard GR case in the top-left panel, the Hernquist-type (top-right panel)  or  NFW-type (bottom-left panel) dark matter model and sGB theory,  the vacuum and two dark matter models of Hernquist-type/NFW-type in the bottom-right panel.  The scalar charges in sGB theory are set to $q_s=\{0.01,0.05,0.08,0.1,0.3,0.5\}$.
}\label{fig:dephasing}
\end{figure*}

\begin{figure*}[htb!]
\centering
\includegraphics[width=7.2in, height=3.2in]{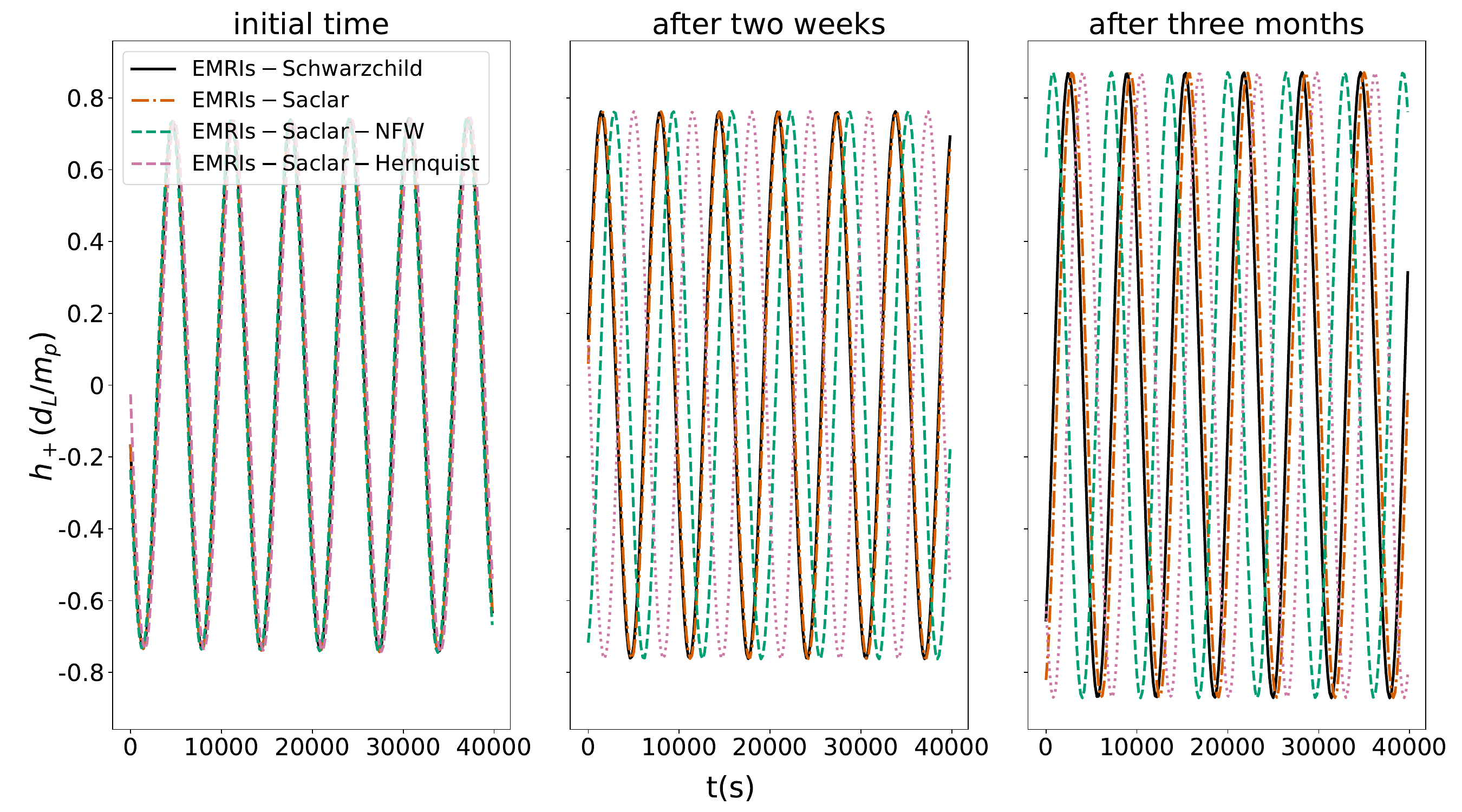}
\caption{
Comparison of the plus polarization $h_{+}$ of EMRI waveforms between the standard GR and dark matter environments, setting scalar charge $q_s= 0.01$. The time-domain waveforms display four cases: signal from Schwarzschild spacetime, including scalar radiation in vacuum and beyond-vacuum of dark matter environments. 
The three panels consider different stages in the time-domain, namely initial time (left panel), the evolutions of two weeks  (middle panel) and three months (right panel).
Inclusion of dark matter environments  can yield phase shifts and amplitude modulations relative to the GR case, highlighting the impact of the environments on the waveform morphology.
}\label{fig:wave}
\end{figure*}

\begin{figure*}[htb!]
\centering
\includegraphics[width=7.3in, height=3.2in]{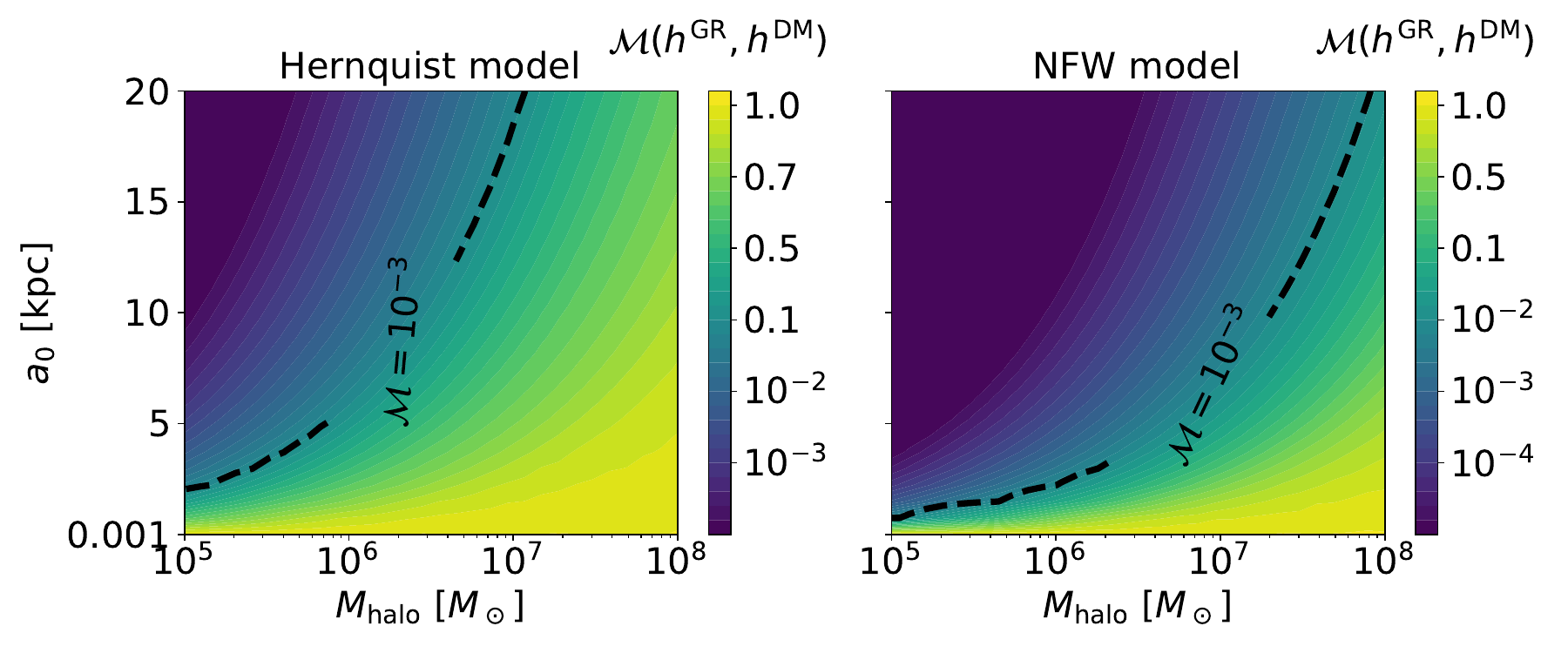}
\caption{
Mismatch as a function of the length scale $a_0$ and halo mass $\Mh$ for the Hernquist (left panel) and NFW (right panel) dark matter profiles. The secondary is initialized at $r_{\Omega} = 10\Mbh$ and evolved adiabatically down to the Schwarzschild ISCO. The black dashed line marks the threshold mismatch $\mathcal{M} = 10^{-3}$, above which the two waveforms are expected to be distinguishable by LISA detector.}\label{fig:mismatch}
\end{figure*}

\begin{figure*}[htb!]
\centering
\includegraphics[width=7.2in, height=4.5in]{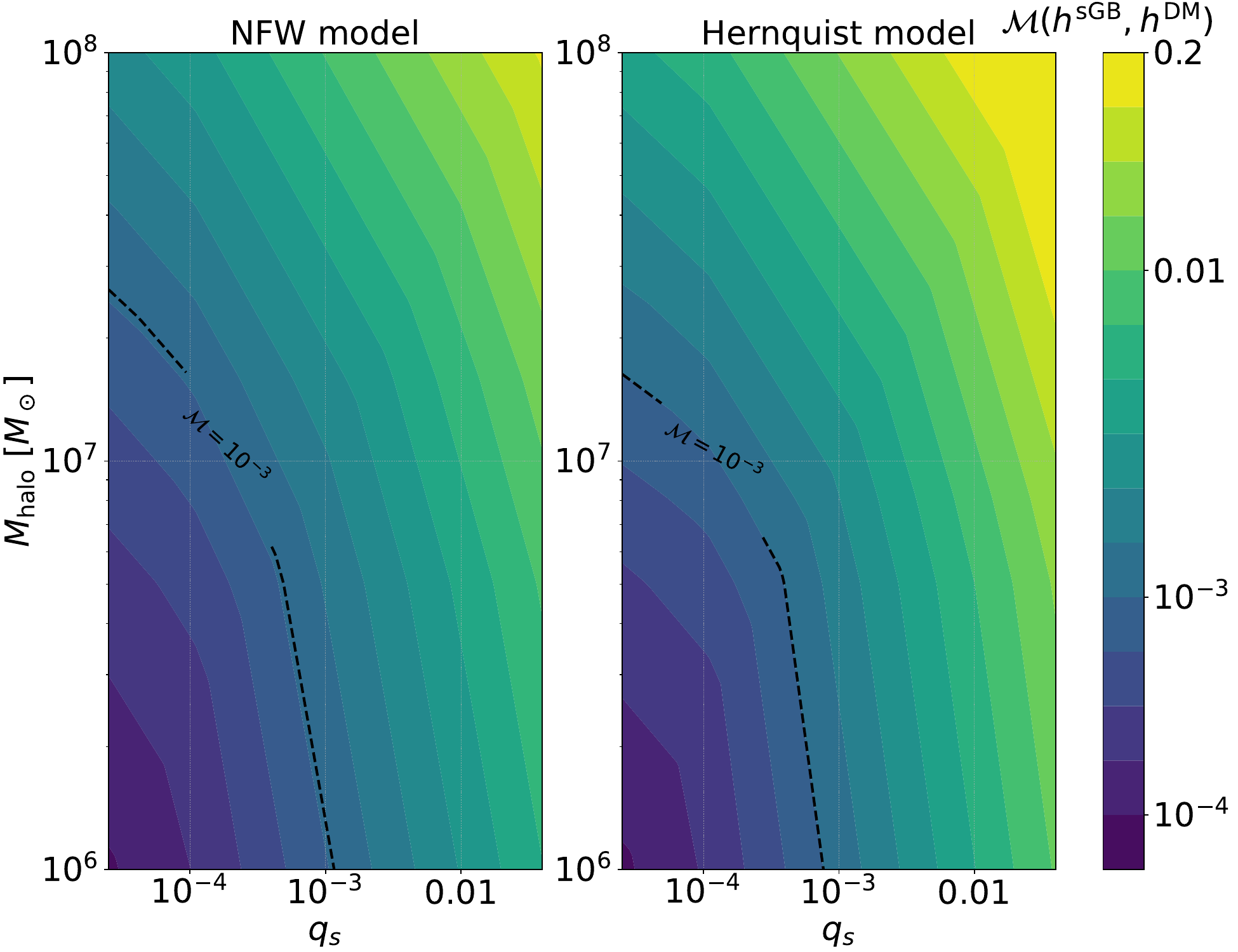}
\caption{
Mismatch between EMRI waveforms in vacuum and in the dark matter environment as a function of the scalar charge and halo mass, for the NFW-type (left panel) and Hernquist-type (right panel) dark matter density profiles. In both cases, the halo scale radius is fixed to $a_0 = 10~\mathrm{kpc}$. The inspiral is initialized at $r_{\Omega} = 10\Mbh$ and evolved adiabatically down to the ISCO of Schwarzschild. The black dashed line indicates the threshold mismatch $\mathcal{M} = 10^{-3}$, above which the two waveforms are expected to be distinguishable by LISA-like detectors.
}\label{fig:mismatch2}
\end{figure*}

\subsection{Inspirals modified by scalar radiation and dark matter environments}
In this subsection, we first present the results of EMRI fluxes incorporating the effects of scalar charge and dark matter environments, then compute the phase difference (dephasing) induced by the environmental effects.
\par
\autoref{fig:flux} shows the behavior of the energy fluxes as functions of the orbital radius $r_{\Omega}/\Mbh$. The left panel shows the ratio of the DM-induced correction to the GW flux relative to its vacuum GR value, $\left|\dot{\textup{E}}^{(1,1)}_{\textrm{GW}}/\dot{\textup{E}}^{(1,0)}_{\textrm{GW}}\right|$, for the Hernquist and NFW type dark matter spike profiles. In this calculation, the numerical parameters are taken to be the same as those presented in \autoref{tab:Fitting_Parameters}, namely $\Mh = 10^{4}\Mbh$, $a_{0} = 10^{3}\Mh$, and $r_{c} = 100\Mh a_{0}/\Mbh$, with the fitting parameters $(\mfa\,,\mfb\,,\mfc)$ also taken from same table. As shown in the figure, in the strong-gravity regime the DM-induced corrections are approximately one order of magnitude smaller than the corresponding vacuum contribution. Moreover, the flux corrections in the NFW case are smaller than those for the Hernquist profile. This behavior can be attributed to the ratio of the corresponding dark matter parameters, $\xi_{\textrm{Hern}}:\xi_{\textrm{NFW}} = 5.65\times 10^{-4}:1.53 \times 10^{-4}\approx 3.63$, for the numerical values adopted in \autoref{tab:Fitting_Parameters}. Since we consider the leading-order correction (linear in $\xi$) to the GW flux due to the presence of dark matter, the correction for the Hernquist profile is expected to be approximately $3.63$ times larger than that for the NFW profile, in agreement with the behavior observed in the plot. Consequently, the NFW-type spike profile has a weaker impact on the corrections to the orbital evolution compared to the Hernquist profile. For both models, $\left|\dot{\textup{E}}^{(1,1)}_{\textrm{GW}}/\dot{\textup{E}}^{(1,0)}_{\textrm{GW}}\right|$ increases monotonically as the orbital radius decreases down to a certain value $r_{\Omega} \approx 7\Mbh$, where we observe a local maximum. This behavior is independent of the specific dark matter profile. The correction then decreases as the orbit approaches the ISCO. Note that, the feature has been reported in Ref.~\cite{Gliorio:2025cbh}. 
\par
In the right panel of \autoref{fig:flux}, we show the ratio between scalar  and gravitational wave flux $\left|\dot{\textup{E}}^{(1,0)}_{\textrm{SW}}/\dot{\textup{E}}^{(1,0)}_{\textrm{GW}}\right|$ for different values of the scalar charge. Consistent with Refs.~\cite{Maselli:2020zgv,Maselli:2021men,Barsanti:2024kul}, for a fixed orbital configuration, the scalar flux is only related to the dimensionless scalar charge. In our setup, the scalar flux increases as the secondary inspirals toward the MBH and approaches the ISCO. However, the ratio $\left|\dot{\textup{E}}^{(1,0)}_{\textrm{SW}}/\dot{\textup{E}}^{(1,0)}_{\textrm{GW}}\right|$ decreases at small separations because the scalar flux grows more slowly than the GW flux in the near-ISCO region. Overall, the relative difference between the total flux in GR and in the sGB theory is at the level of \(\sim 2\%\) close to the plunge.

As shown in \autoref{fig:flux}, the flux ratios in the presence of dark matter and scalar charge provide an estimate of how the surrounding environment and sGB corrections influence the waveform of EMRIs. To assess these effects more precisely, we compute the accumulated phase difference, or dephasing, between the perturbed and reference waveforms.
\autoref{fig:dephasing} displays the dephasing as a function of the observation time for various configurations of the DM density profile and scalar charge in the sGB framework.

To disentangle the influence of different environmental configurations on EMRIs orbital dynamics, we compare the dephasing between (i) GR and sGB gravity, (ii) Hernquist-type or NFW-type dark matter environments and sGB gravity, and (iii) the two dark matter environments and the vacuum (Schwarzschild) spacetime, as shown in \autoref{fig:dephasing}. Across the three panels (the top two and bottom-left), the dephasing is consistently correlated with the magnitude of the scalar charge. For the comparison between GR and sGB gravity, the threshold scalar charge distinguishable by LISA is approximately $q_s = 0.01$. However, the same minimum value $q_s = 0.01$ is not easily distinguishable for either the Hernquist-type or NFW-type dark matter profiles.

When contrasting the dark matter environments with the Schwarzschild vacuum, we find that the accumulated dephasing can reach $\mathcal{O}(10^{4})$ radians with one-year observation of LISA. This indicates that both the presence of dark matter and the scalar charge can imprint observable signatures on the EMRIs inspiral waveform. From the dephasing analysis, we can therefore infer that these environmental and beyond-GR effects may be detectable within LISA’s sensitivity range. To quantitatively evaluate the degree of distinguishability among different environmental configurations, we further compute the waveform mismatches of EMRI signals under various settings; these results are presented in \autoref{wave:discrimination}.

\subsection{EMRI waveforms and their discrimination}\label{wave:discrimination}
In this subsection, we compare the EMRI waveforms in vacuum and dark matter environments. \autoref{fig:wave} shows the plus polarization of the waveform, rescaled by a factor of $d_L/m_p$, for four representative cases: the pure GR (vacuum) case, scalar radiation in vacuum, and the dark matter environments described by the NFW and Hernquist profiles.

As shown in the left panel of \autoref{fig:wave}, the phases of all four waveforms remain nearly identical during the initial $4\times10^4~\mathrm{s}$ of inspiral, indicating that the influence of scalar radiation and dark matter is negligible at early stages. After approximately two weeks of evolution (middle panel), a clear phase deviation emerges among the waveforms. After three months of inspiral, all cases exhibit distinct phase offsets relative to the GR reference. These results suggest that the presence of dark matter halos induces a cumulative dephasing effect in the EMRI signals, which becomes significant over long observation times.

To assess the influence of dark matter spikes on the detectability of environmental effects in EMRIs, we compute the waveform mismatches between Schwarzschild and dark matter-immersed systems, as illustrated in \autoref{fig:mismatch} and \autoref{fig:mismatch2}. The mismatch is evaluated as a function of both the total dark matter spike's mass and the scale radius, considering two density profiles: Hernquist and NFW model. Three distinct EMRI waveforms are analyzed: $h^{\rm DM}$, corresponding to the inspiral within a dark matter environment; $h^{\rm sGB}$, representing the waveform predicted by sGB gravity; and $h^{\rm GR}$, describing the inspiral in a Schwarzschild spacetime under GR. We model an EMRI system with component masses $(10^6 + 10)\,M_\odot$, where the secondary compact object, endowed with a scalar charge $q_s = 0.01$, begins its inspiral from $r_{\Omega} = 10\,M_{\rm BH}$ within the dark matter halo. The impact of the dark matter environment on the EMRI waveform is quantified through the mismatches $\mathcal{M}(h^{\rm GR}, h^{\rm DM})$ between the GR and DM cases, and $\mathcal{M}(h^{\rm sGB}, h^{\rm DM})$ between the sGB and DM scenarios. 

\autoref{fig:mismatch} illustrates how variations in the halo parameters affect the mismatch $\mathcal{M}(h^{\rm GR}, h^{\rm DM})$. For a fixed halo mass, a smaller scale radius leads to a larger mismatch, since a more compact halo leads to larger values of dark matter parameter $\xi$, which in turn  leads to a stronger gravitational perturbation on the orbital motion. Conversely, increasing the scale radius renders the halo more diffuse, reducing its effect on the waveform. In both models, more massive dark matter halos produce mismatches exceeding the threshold value of $10^{-3}$, implying that such environmental effects could be observable by LISA-like detectors. Moreover, for the same halo mass and scale radius, the Hernquist profile yields systematically larger mismatches than the NFW profile, consistent with the findings of Ref.~\cite{Zhang:2024ugv}.

We further analyze the dependence of the mismatch 
$\mathcal{M}(h^{\rm sGB}, h^{\rm DM})$ on the scalar charge and dark matter-halo mass for a fixed length scale $a_0 = 10~\mathrm{kpc}$ of dark matter spike, as presented in \autoref{fig:mismatch2}. The mismatch between the vacuum and non-vacuum waveforms is shown for different halo masses $\Mh$ and scalar charges $q_s$, assuming that the secondary begins its inspiral from $r_{\Omega} = 10\Mbh$ and evolves adiabatically down to the Schwarzschild ISCO. For a given halo mass, the mismatch increases with $q_s$, reflecting the quadratic dependence of the scalar flux on the charge. Similarly, more massive halos enhance the mismatch, confirming that both scalar and dark matter effects imprint measurable signatures on EMRI waveforms.

\begin{table*}[htbp!]
\centering
\begin{tabular}{cc|ccccccccccc}
\hline
\hline  
$\rm DM ~profile$ & $q_s$  & $\sigma_{\Mbh}/\Mbh$ & $\sigma_{m_p}/m_p$   &$\sigma_{\mfa}$ & $\sigma_{\mfb}$ & $\sigma_{\mfc}$   
&$\sigma_{a_0}/a_0$   &$\sigma_{\Mh}/\Mh$ 
& $\sigma_{q_s}/q_s$   &$\sigma_{r_{\Omega}}$
& $\sigma_{\phi{_{p,0}}}/\phi_{p,0}$
\\
\hline
$\rm Hernquist $  & $0.01$
&$5.09\text{e-5}$    &$4.84\text{e-4}$   &$4.27\text{e-3}$ 
&$3.58\text{e-3}$    & $2.16\text{e-3}$
&$2.13\text{e-2}$   &$5.04\text{e-2}$ 
&$3.20\text{e-2}$  &$8.26\text{e-4}$  &$5.36\text{e-1}$  
\\
& $0.1$
&$4.25\text{e-5}$    &$3.24\text{e-4}$   &$3.56\text{e-3}$ 
&$3.28\text{e-3}$    & $2.45\text{e-3}$
&$3.46\text{e-2}$   &$4.37\text{e-2}$ 
&$3.43\text{e-2}$  &$5.57\text{e-4}$  &$4.75\text{e-1}$  
\\
\hline
$\rm NFW $  & $0.01$
&$5.86\text{e-5}$    &$4.29\text{e-4}$   &$5.46\text{e-3}$ 
&$5.91\text{e-3}$    & $3.69\text{e-3}$
&$7.56\text{e-2}$   &$5.12\text{e-2}$ 
&$6.74\text{e-2}$  &$9.31\text{e-4}$   &$7.06\text{e-1}$  
\\
& $0.1$
& $4.72\text{e-5}$    & $3.64\text{e-4}$   & $4.12\text{e-3}$ 
& $4.56\text{e-3}$    & $3.35\text{e-3}$
& $5.27\text{e-2}$   & $4.37\text{e-2}$ 
& $4.61\text{e-2}$  & $3.53\text{e-4}$  & $5.27\text{e-1}$  
\\
\hline
\hline
\end{tabular}
\caption{Measurement uncertainties for intrinsic EMRI parameters in non-vacuum environments. 
We consider binaries with component masses $(\Mbh=10^{6}M_\odot,\; m_p=10M_\odot)$, scalar charge $q_s\in[0.01,0.1]$, and initial periapsis $r_{\Omega}=10M$. 
The inspiral is evolved for two years; the luminosity distance $d_L$ is chosen such that the network SNR is $50$. 
Waveforms are generated for two dark matter profiles with spike–shape fit parameters $(\mfa,\mfb,\mfc)$, taking $(\mfa,\mfb,\mfc)=(2.64,\,2.33,\,0.544)$ for NFW and $(\mfa,\mfb,\mfc)=(2.64,\,2.33,\,1.34)$ for Hernquist. 
The macroscopic parameters of dark matter  halo are fixed to the setting of \autoref{tab:Fitting_Parameters}. 
The initial orbital phase is $\phi_{p,0}=1.0$, and the source–detector geometry is set by $(\theta_S,\phi_S,\theta_K,\phi_K)=(1.0,1.0,1.0,1.0)$.}\label{tab:fim:error}
\end{table*}

\begin{figure*}[htb!]
\centering
\includegraphics[width=0.87\paperwidth]{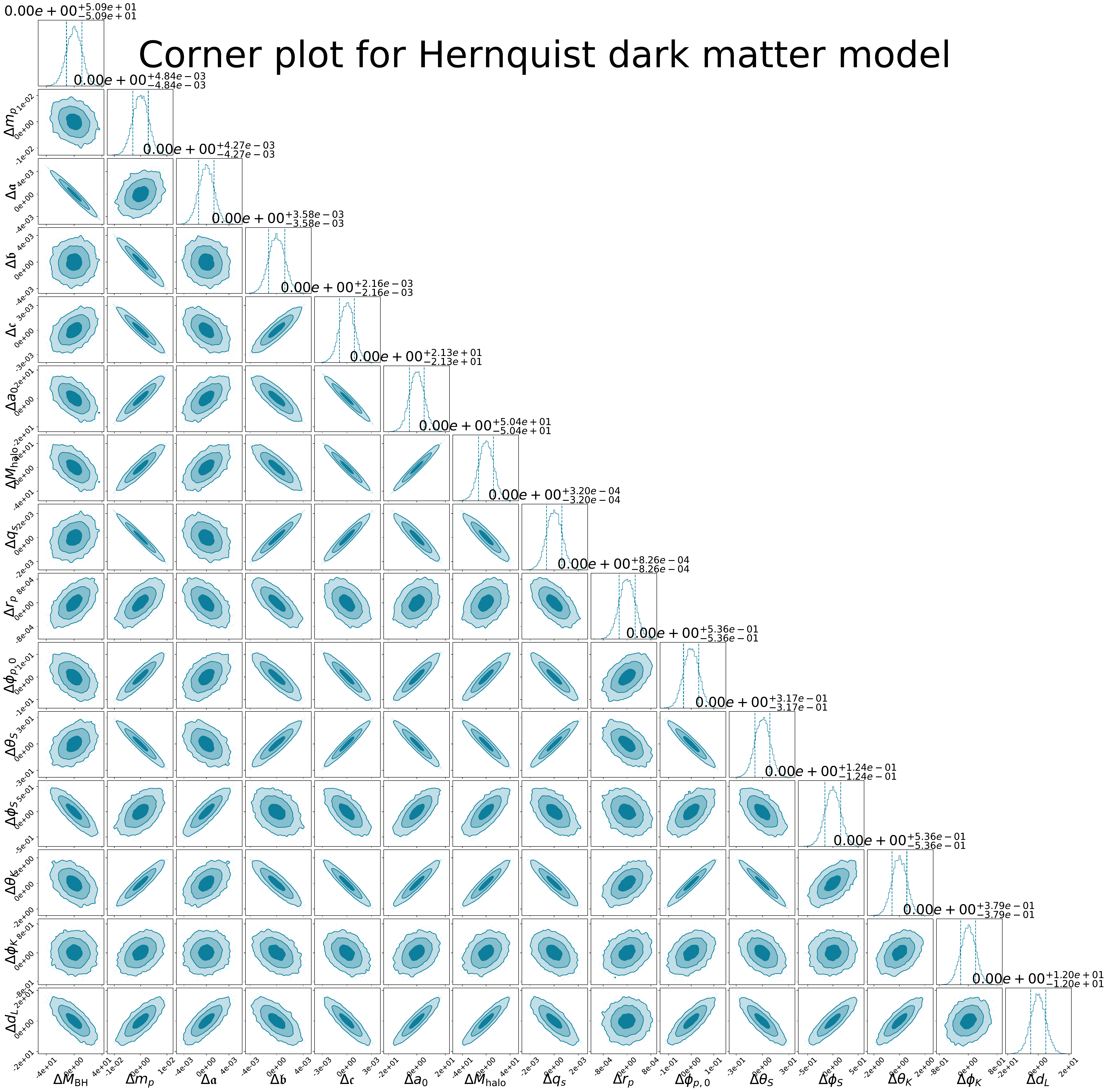}
\caption{Corner plot of posterior probability densities for EMRIs embedded in a Hernquist-type dark matter environment. We assume $(\Mbh=10^{6}M_\odot,\; m_p=10M_\odot,\; r_{\Omega}=10\Mbh,\; \phi_{p,0}=1.0,\; q_s=0.01,\;)$ and the macroscopic parameters of Hernquist-type dark matter halo are fixed to the setting of \autoref{tab:Fitting_Parameters}. All extrinsic parameters are fixed as in \autoref{tab:fim:error}. 
Posteriors are inferred from a two-year LISA observation. 
Vertical dashed lines denote the $1\sigma$ credible intervals;
shaded contours show the $68\%$, $95\%$, and $99\%$ credible regions.}\label{fig:cornerplot:Hern}
\end{figure*}

\begin{figure*}[htb!]
\centering
\includegraphics[width=0.87\paperwidth]{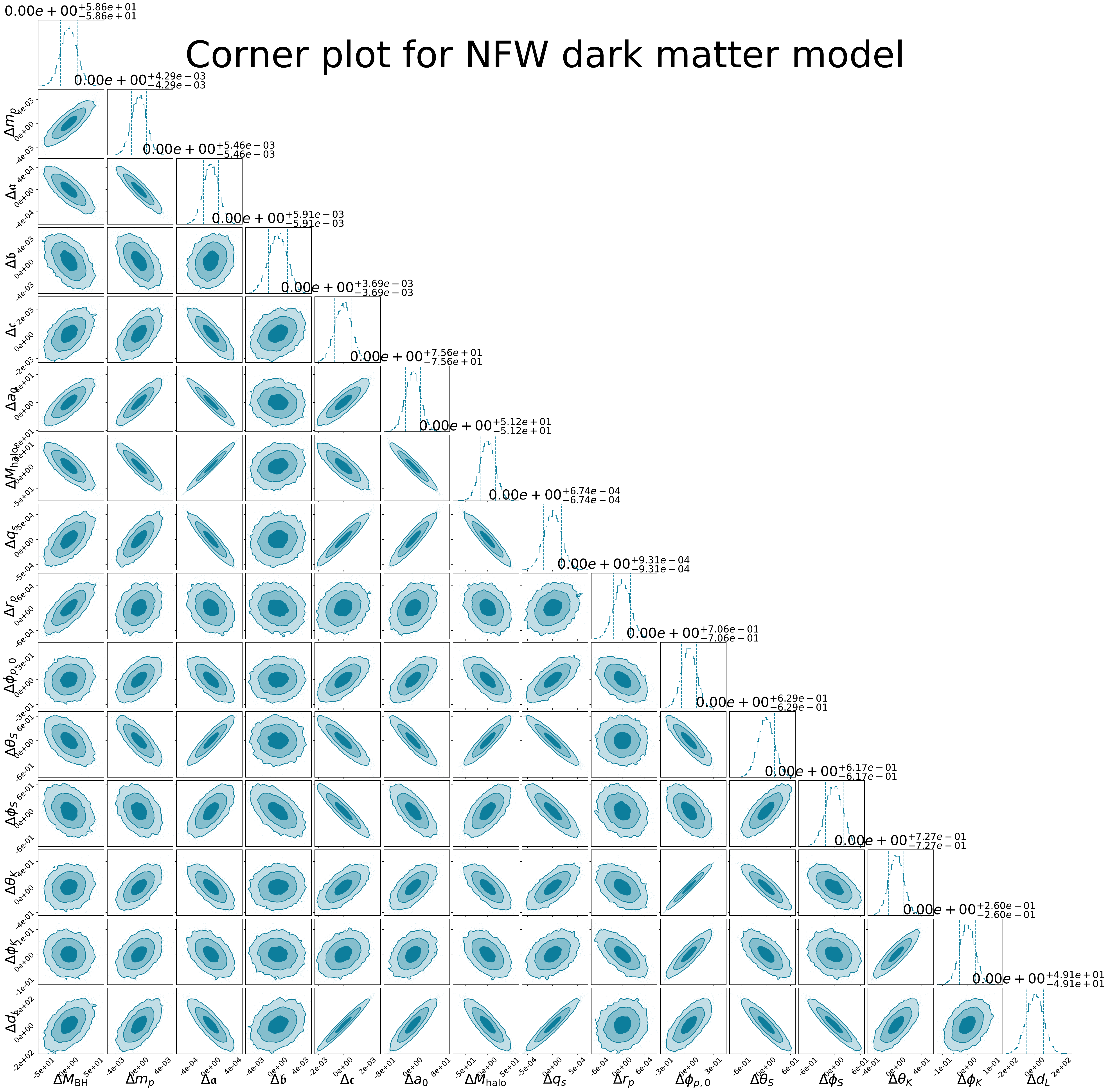}
\caption{Corner plot of posterior distributions for EMRIs embedded in an NFW-type dark matter environment. We adopt $(\Mbh=10^{6}M_\odot,\; m_p=10M_\odot,\; r_{\Omega}=10\Mbh,\; \phi_{p,0}=1.0,\; q_s=0.01,\;)$ and the macroscopic parameters of Hernquist-type dark matter halo are set to the configuration of \autoref{tab:Fitting_Parameters}, with all extrinsic parameters fixed as in \autoref{tab:fim:error}. 
Posteriors are inferred from a two-year LISA observation. 
Vertical dashed lines mark the $1\sigma$ credible intervals, and shaded contours denote the $68\%$, $95\%$, and $99\%$ credible regions.}\label{fig:cornerplot:NFW}
\end{figure*}

\begin{figure*}[htb!]
\centering
\includegraphics[width=4.17in, height=3.13in]{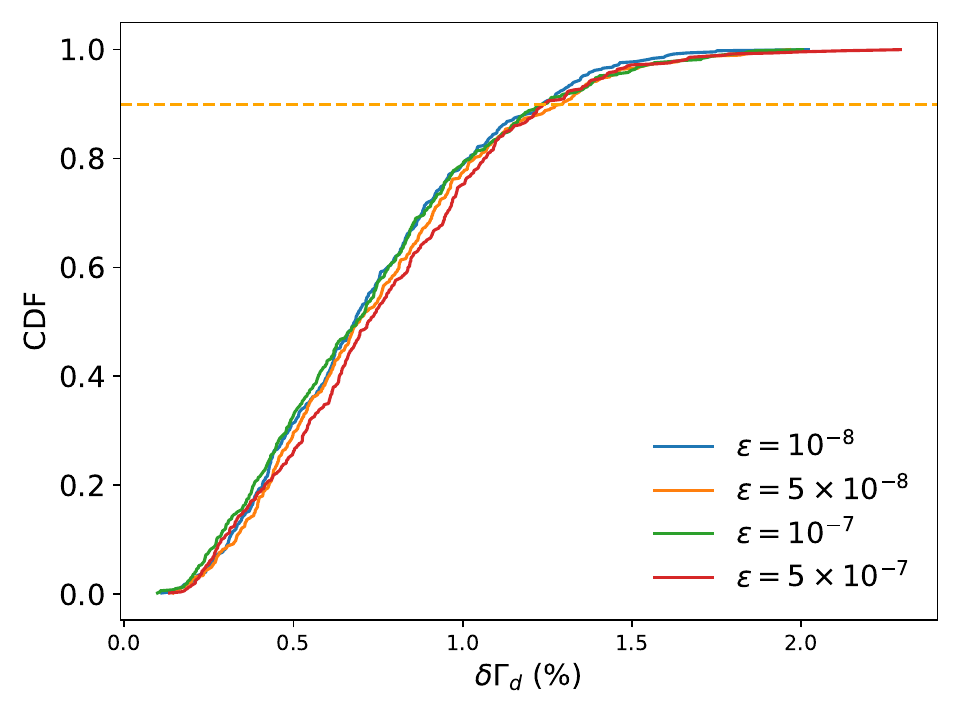}
\caption{Cumulative distributions of the maximum relative discrepancy between the Fisher matrix and a perturbed version, where each element of the perturbation matrix is drawn from a uniform distribution $u \in [-10^{-3},10^{-3}]$. 
The colored curves correspond to different choices of the finite-difference spacing, $\epsilon \in \{10^{-8},\,5\times10^{-8},\,10^{-7},\,5\times10^{-7}\}$, employed in the numerical evaluation of the FIM. The example shown focuses on derivatives with respect to the dark matter halo's parameter $a_0$. 
All other parameter choices in the two panels are identical to those adopted in \autoref{fig:cornerplot:Hern}.}
\label{fig:CDF:plot}
\end{figure*}

\subsection{Constraint on dark matter environments}
Here we constrain the scalar charge and dark matter halo parameters using EMRI signals in LISA-like detectors within the low-frequency approximation. Because fully Bayesian inference with Markov-Chain Monte Carlo would require millisecond-scale waveform generation and substantial computational resources, we instead employ the FIM formalism to estimate uncertainties for the parameter vector in \autoref{eq:parameters:vector}.

For a specified dark matter profile (NFW or Hernquist), we compute the inspiral by incorporating the dark matter and scalar-induced modifications to the fluxes, construct the quadrupolar waveform via \autoref{amplitude}, and project it onto the detector response using \autoref{antenna}. Although the FIM is evaluated over all parameters in \autoref{eq:parameters:vector}, we report in \autoref{tab:fim:error} only the uncertainties for intrinsic parameters, since the EMRIs phase is largely insensitive to extrinsic parameters whose effects are predominantly absorbed by overall amplitude and orientation.

Under these assumptions, the fractional uncertainties for the dark matter spike shape parameters $(\mfa,\mfb,\mfc)$ are at the level of \(\sim10^{-3}\), while the characteristic scale \(a_0\) and the total halo mass \(\Mh\) are measured to \(\sim10^{-2}\). Comparing profiles, the Hernquist model is constrained more tightly than NFW, consistent with its steeper inner slope (larger \(\mfc\)) which imprints a stronger phase modulation. For EMRIs embedded in dark matter environments, the scalar charge is constrained at the \(\sim10^{-2}\) level, with the Hernquist profile again yielding slightly stronger bounds than NFW.

According to the FIM method, the $1\sigma$ uncertainty of each parameter is inferred from the diagonal elements of the covariance matrix, which is defined as the inverse of the Fisher matrix. The covariance matrix can also be used to assess correlations among source parameters, which in turn helps to clarify the relationships relevant to detecting the scalar charge in the non-vacuum case.
As shown in \autoref{fig:cornerplot:Hern} and \autoref{fig:cornerplot:NFW}, distinct correlation patterns appear across the full parameter space. Specifically, there is a positive correlation between the scalar charge $q_s$ and the parameters $(\mfb,\mfc)$ for the Hernquist-type dark matter model, whereas the correlation between $q_s$ and $\mfb$ is weak for the NFW-type case. The correlation between $q_s$ and the halo scale $a_0$ changes from positive to negative when moving from the Hernquist-type to the NFW-type model, while an opposite trend is observed for the correlation between $q_s$ and the secondary object's mass $m_p$. Moreover, the correlations between $q_s$ and the dark matter-halo mass $\Mh$ are negative for both dark matter models.
These correlation analysis indicate that the accurate measurements of intrinsic parameters are very vital to put a tighter bounds on scalar charge in the beyond-vacuum and break the potential degeneracies in the parameter estimation of EMRIs.
Additionally, one can find that the correlation between $q_s$ and other extrinsic parameters 
$(\theta_{S,K}, \phi_{S,K}, d_L)$ do not still keep invariant for two dark matter models.
Therefore, we should deal with the fluxes and waveforms of EMRIs beyond-vacuum carefully to place a moderate constraint on scalar charge from the modified gravities.

As discussed in previous subsections, the measurement uncertainty of each source parameter inferred from the FIM strongly depends on the numerical stability of the covariance matrix. To assess this, we evaluate the robustness of the FIM and its inversion following Refs.~\cite{Speri:2021psr,Maselli:2021men,Zi:2022hcc}. We first construct a perturbation matrix $\mathbf{U}$ with the same dimension as the FIM $\mathbf{\Gamma}$, whose elements are randomly drawn from a uniform distribution $u \in [-10^{-3},10^{-3}]$. We then compute the inverse of the perturbed matrix $(\mathbf{\Gamma}+\mathbf{U})$ and evaluate the maximum relative deviation between the perturbed and unperturbed covariance matrices, defined as
\begin{equation}
\delta \mathbf{\Gamma}_d \equiv \max \left( \frac{(\mathbf{U}+\mathbf{\Gamma})^{-1}-\mathbf{\Gamma}^{-1}}{\mathbf{\Gamma}^{-1}} \right).
\end{equation}

We compute the numerical stability of the covariance matrix using the EMRI waveforms from the environments of Hernquist-type dark matter halo in \autoref{fig:CDF:plot}, where the numerical derivation of waveforms with respect to the parameter $a$ of the dark matter model is set to $\epsilon\in\{10^{-8},5\times10^{-8},10^{-7},5\times10^{-7}\}$. From the figure of the cumulative distribution function, one can find that the distribution exhibits good numerical stability. Specifically, more than $90\%$ of the population have the relatively smaller values below $1.2\%$, indicating that the maximum error lies safely in the tolerance needed for computing a stable Fisher-matrix inversion.
Here we only have the case of the numerical differential interval for the parameter $a$ of the dark matter halo, there also exists a similar conclusion for the case of parameters $(\mfb,\mfc,a_0,\Mh)$ of dark matter halo, so we do plan to show the full results for those parameters and the model of NFW-type dark matter.

\section{Conclusion}\label{secVI}
EMRIs offer a powerful way to study gravity/GWs in the strong-field regimes. Because of their long-lived signals and large accumulated phase, EMRIs are sensitive not only to the vacuum dynamics for GR but also to deviations caused by surrounding matter and possible extensions of GR. As black holes are expected to reside in environments that can potentially impact the GW radiation and inspiral dynamics, in parallel, modified gravity theories like sGB give rise to additional radiation (scalar radiation). When such effects are simultaneously present, they can lead to systematic biases in parameter estimation. Therefore, disentangling them is crucial for extracting reliable constraints on beyond-vacuum physics from future GW observations.

In this paper, following the method of Refs.~\cite{Sadeghian:2013laa,Rahman:2025mip}, we consider the more realistic astrophysical configuration that EMRI evolution is influenced by the dark matter density profiles, where both Hernquist and NFW profiles are included. Assuming that dark matter around MBH as the anisotropic fluid with a zero radial pressure, we compute the modification of dark matter on EMRIs fluxes at the first post-adiabatic order. In the sGB theory, using the method developed in Refs.~\cite{Maselli:2020zgv,Barsanti:2022gjv,Xie:2024xex,DellaRocca:2024sda}, the spacetime geometry of MBH can be described by the Schwarzschild metric because the scalar charge carried by MBH is suppressed by the magnitude of the curvature around the horizon, whereas the secondary object carries a non-negligible scalar charge.
Therefore, we can compute the scalar fluxes under the Regge-Wheeler-Zerilli framework to obtain the total energy fluxes, then evolve the orbital parameters for the circular EMRIs on the equatorial plane.

Notably, we compute the gravitational energy fluxes for both Hernquist and NFW dark matter profiles and compare their behavior at the adiabatic and first post-adiabatic orders. Our analysis demonstrates that the gravitational radiation overwhelmingly dominates the inspiral evolution, while the scalar emission contributes only subdominantly to the overall energy loss. At post-adiabatic order, the influence of dark matter is found to be stronger for the Hernquist profile than for the NFW-type distribution, indicating a more pronounced modification to the inspiral dynamics in the former case.

Regarding the phase evolution, we find that the presence of dark matter induces a significant cumulative dephasing of approximately $\Delta \Phi^{\rm DM, vac} \sim \mathcal{O}(10^{4})$ radians for both the Hernquist-type and NFW-type profiles, relative to the vacuum scenario. In sGB gravity, the scalar charge carried by the secondary object produces a dephasing of $\Delta \Phi^{\rm GR, sGB} \sim \mathcal{O}(800)$ radians for a representative scalar charge of $q_s = 0.5$. When both dark matter and sGB effects are simultaneously considered, the total dephasing reaches approximately $\Delta \Phi^{\rm DM, sGB} \sim \mathcal{O}(10^{3})$ radians for the same scalar charge. Furthermore, the dephasing induced by the scalar charge is found to be highly sensitive to its magnitude, increasing monotonically with larger values of $q_s$.
To quantitatively assess the impact of dark matter on EMRI waveforms, we compute the waveform mismatches under different environmental configurations, namely: (i) the Schwarzschild spacetime versus dark matter environments modeled by the Hernquist-type and NFW-type profiles, and (ii) the Schwarzschild spacetime versus the sGB scenario. Our mismatch analysis shows that, for EMRI signals,  LISA would be capable of distinguishing the effects of dark matter when the halo is more massive and characterized by a smaller length scale. In particular, compared to the NFW-type model, the Hernquist profile exhibits a broader distinguishable parameter space in $(\Mh, a_0)$.
When comparing EMRI waveforms between the Schwarzschild spacetime and the sGB gravity scenario within dark matter environments, we find that the mismatch threshold depends sensitively on the scalar charge. Specifically, for the NFW-type profile, configurations with $(q_s \gtrsim 10^{-3}, \Mh\gtrsim 10^{7.4} M_\odot)$ at a fixed dark matter scale radius of $a_0 = 10~{\rm kpc}$ are distinguishable by LISA. Under the same conditions, the Hernquist profile yields a slightly lower distinguishable threshold of $\Mh \gtrsim 10^{7.2} M_\odot$.
Finally, we constrain the scalar charge and dark matter parameters using the FIM analysis. For both NFW- and Hernquist-type dark matter models, LISA can measure the scalar charge with a fractional uncertainty of order $\sim 10^{-2}$, with a marginally better precision achieved for the Hernquist profile due to its stronger dark matter-induced modulation of the waveform.

In our current analysis, several simplifying assumptions have been adopted. First, the computation of the EMRI waveforms is based on the quadrupole formula, rather than the fully relativistic waveform derived from black hole perturbation theory. Second, the modeling of the adiabatic inspiral relies on interpolated energy fluxes; therefore, a detailed assessment of various interpolation schemes is necessary to evaluate their feasibility for detecting signatures of non-GR effects in beyond-vacuum environments~\cite{Khalvati:2025znb}. Third, since MBHs in the universe are generally expected to possess high spins, it is essential to extend our analysis to include the influence of dark matter environments within both GR and alternative gravity frameworks using relativistic perturbation theory. We anticipate that the present study provides a foundational step toward accurately modeling the impact of rotating MBHs on EMRI dynamics and GW signals.

In addition, this work opens up several other new and interesting avenues for further exploration. It enables us to examine the beyond-vacuum scenario with secondary exhibiting eccentric orbits as well as eccentric-inclined orbits in future work \cite{Fujita:2009us, Piovano:2024yks, Zi:2025lio}. This paves the way for a natural extension and more astrophysically realistic setting of the present work. We note that the EMRI signal response adopted here relies on low-frequency considerations; consequently, extending the analysis to include time-delay interferometry is an important avenue for improving the inferred constraints \cite{Tinto:2004wu, Tinto:2022zmf, Wu:2023key}. As constraints on beyond-vacuum GR presented here are derived using the FIM, the Bayesian Markov Chain Monte Carlo (MCMC) treatment may yield more stringent results \cite{Katz:2021yft, Chua:2021aah}. Moreover, extending the present work to population-based EMRIs will improve the statistical robustness and allow for additional investigation of how beyond-vacuum effects accumulate across numerous sources \cite{Kejriwal:2025jao}. Furthermore, including additional modified gravity theories, including cases where the background spacetime deviates from Schwarzschild geometry, contrary to the sGB framework presented here, together with environmental contributions, would generalize the analysis further and enable a quantitative assessment of biases arising from unaccounted non-vacuum effects \cite{Julie:2019sab, Sotiriou:2013qea}. Apart from these, the modified Teukolsky framework will also be an interesting approach to investigate such effects and will serve as an consistency check and robustness of the method for precision waveform modelling \cite{Li:2022pcy, Li:2025ffh, Kumar:2025jsi}. We aim to investigate some of these studies in our future work.

\section{Acknowledgements}
T. Zi. is funded by the National Natural Science Foundation of China with Grants No. 12347140 and No. 12405059, key Program of the Natural Science Foundation of Jiangxi Province under Grant No. 20232ACB201008, and the Ganpo High-Level Innovative Talent Program.  M.~R. is partially supported by the JSPS KAKENHI Grant No.~ JP23KF0233. S. K. is funded by National Post-Doctoral Fellowship (Grant No. PDF/2023/000369) from the Anusandhan National Research Foundation  (ANRF, formerly SERB), Department of Science and Technology (DST), Government of India.

\bibliography{0_ref}
\bibliographystyle{utphys1}
\end{document}